\documentclass[aps,prl,reprint,superscriptaddress]{revtex4-2}
\usepackage{verbatim}
\usepackage{graphicx}
\usepackage[export]{adjustbox}
\usepackage{float}
\usepackage{amssymb}
\usepackage[utf8]{inputenc}
\usepackage[english]{babel}
\usepackage{amsmath}
\usepackage{mathtools}
\usepackage{amsfonts,xcolor}
\usepackage{stackrel}
\usepackage{cases}
\usepackage{bm}%
\usepackage{hyperref}
\usepackage{tikz}
\usepackage{pgfplots}
\usepgfplotslibrary{fillbetween}
\pgfplotsset{compat=1.18}
\usetikzlibrary{calc}
\usetikzlibrary{decorations.pathreplacing}
\usetikzlibrary{decorations}
\usetikzlibrary{arrows,decorations.markings}

\newcommand{\eps}{\varepsilon}

\newcommand{\im}{\mathbf{i}}
\newcommand{\be}{\begin{align}}
\newcommand{\ee}{\end{align}}
\newcommand{\C}{\mathbb{C}}
\newcommand{\D}{\mathbb{D}}

\newcommand{\R}{\mathbb{R}}
\newcommand{\E}{\mathbb{E}}
\newcommand{\N}{\mathbb{N}}

\usepackage{color}
\usepackage{verbatim}

\definecolor{blue}{rgb}{0,0,1}
\definecolor{green}{rgb}{0,.75,0}
\definecolor{red}{rgb}{0,0,0}
\definecolor{vio}{rgb}{1,0,1}
\definecolor{uv}{rgb}{0.5,0,0.5}
\definecolor{ama}{rgb}{0.3,0.3,0.3}

\begin{document}
\title{Probabilistic construction of non compactified imaginary Liouville field theory}
\author{Romain Usciati}\affiliation{LPTMS, CNRS, Univ. Paris-Sud, Université Paris-Saclay, 91405 Orsay, France}
\author{Colin Guillarmou}\affiliation{IMO, Universit\'e Paris-Saclay, France}\author{Remi Rhodes}\affiliation{Aix Marseille Universit\'e, CNRS, I2M, Marseille, France
}\author{Raoul Santachiara}\affiliation{LPTMS, CNRS, Univ. Paris-Sud, Université Paris-Saclay, 91405 Orsay, France}
 \begin{abstract}
  We propose a probabilistic construction of imaginary Liouville Field Theory based on a real (non-compactified) Gaussian Free Field. We argue that our theory is the first explicit Lagrangian field theory that reproduces the imaginary DOZZ structure constants without requiring a neutrality constraint. Our proposal is supported by exact results for the imaginary Gaussian Multiplicative Chaos on the circle, and by numerical simulations on the sphere.  In particular, we show that the three-point functions of the theory agree remarkably well with the imaginary DOZZ structure constants.
 \end{abstract}
 \maketitle 
Over the past decade, mathematicians have developed a probabilistic construction of 2D conformal field theories, thereby providing a rigorous definition of correlation functions in terms of averages  over random Gaussian free fields (GFF)\cite{review24}. 

The Liouville field theory (LFT$_b$) is a one-parameter ($b$) family  of conformal field theories. The regime $b\in \mathbb{R}$, where LFT governs quantum gravity in two dimensions, is very well understood. The greatest achievement of the probabilistic approach is the rigorous derivation of the corresponding bootstrap solutions of LFT$_{b\in \mathbb{R}}$\cite{segal21}.  On the other hand, for pure imaginary values of $b$,  $b=\im \beta$ with $\beta\in \mathbb{R}$ the analytic continuation of the LFT bootstrap solution is impossible. A long-standing question remains: whether and how a probabilistic construction of the LFT$_{\im \beta}$, which we will  refer to  as imaginary LFT, can be achieved. This question naturally arises in the study of geometric statistical models, such as the $O(n)$ loop models \cite{ijs15} and the related Potts models \cite{psvd13,prs16}, and in two-dimensional string theory\cite{Collier_2024}. Indeed, these models have revealed the existence of new conformal field theories, and the investigation of their solutions via the bootstrap approach has seen intense efforts over the past decade, see the recent review \cite{ribault2025} and references therein.

A first step in this direction has been the rigorous definition of a compactified version of the LFT$_{\im \beta}$\cite{guillarmou2023,chatterjee2025}, the compactified imaginary Liouville theory (CILT). This theory, which is based on a GFF taking values on a circle, is strictly related to the Coulomb gas formalism for minimal models \cite{dofa_npb84,nienhuis1987coulomb}. In particular, the correlation functions vanish unless they satisfy a selection rule, known as the neutrality condition. However, these selection rules impose strong constraints, making the theory too limited to describe observables in geometric statistical models or for applications in quantum gravity.

Here, we propose a probabilistic construction of LFT$_{\im \beta}$ based on a real (non-compactified) GFF. We show in particular that our theory is the first explicit Lagrangian field theory that generates the so-called imaginary DOZZ structure constants without requiring a neutrality constraint. The imaginary DOZZ constants govern amplitudes in geometric statistical models, such as the three-point connectivity of random Potts model, and they emerge as a fundamental component of new bootstrap solutions. Our approach establishes a natural extension of compactified imaginary LFT and offers a new perspective on conformal fields theories which have central charge $c_{\text{charge}}\leq 1$, with potential implications for statistical physics and 2D quantum gravity.

Consider a surface $\Sigma$ with boundary $\partial \Sigma$ and equipped with a metric $g$. We denote by ${\rm dv}$ and  ${\rm d}\ell$  the corresponding volume forms on $\Sigma$ and $\partial \Sigma$, and by $K_g$ and $k_g$ respectively the scalar and the geodesic curvature of $\Sigma$ and $\partial \Sigma$. The LFT$_{\im \beta}$ action takes the form:
\begin{align}
\label{imagLFT}
\mathcal{S}^{\text{im}}_{{\rm L}}(\Sigma,g,\phi)=&\frac{1}{4\pi}\int_{\Sigma} \left( |d \phi|^2_{g}+\im Q K_g \phi  + \mu e^{\im \beta \phi}\right){\rm dv} +\nonumber \\
&+\frac{1}{2\pi}\int_{\partial \Sigma} \left(\im Q k_g \phi+\mu_B e^{\im \frac{\beta}{2} \phi}\right){\rm d}\ell,
\end{align}
where $\phi(x)$ is the fundamental field, $\mu,\mu_B \in \mathbb{R}$ are usually known as, respectively, the bulk and the boundary cosmological constants. The parameter $\beta \in \R$ is a real parameter and one sets  $Q=\frac{\beta}{2} - \frac{2}{\beta}$, ensuring that the resulting quantum field theory is conformally invariant with central charge $c_{\text{charge}}=1-6 Q^2 \leq 1$.

Let us discuss first the simplest case of the imaginary LFT$_{\im \beta}$ on a circle. 
The "circle Liouville theory" corresponds to the LFT$_{\im \beta}$ on the flat unit disk, $(\Sigma,g)=(\mathbb{D},g_0=|d z|^2)$  with $\mu=0$. We split the Liouville field $\phi(x)$ as  $\phi(x)= X_{g_0}(x) + c$, where $c$ is a constant and  $X_{g_0}$ is the (real) Neumann GFF on the disk  with covariance $\mathbb{E} [ X_{g_0}(x)X_{g_0}(y)]  =-\log{|x-y||1-x\bar y|}$. As we motivate below, the important difference with respect to previous works is that we allow the constant $c$ to take general complex values, $c\in \mathbb{C}$. 

The basic objects in LFT are defined through an ultraviolet regularization of the GFF $X_{g_0,\epsilon}$ at scale $\epsilon$. The primary field of the imaginary LFT are vertex field $V_{\alpha}$ with scaling dimension $\Delta = \frac{\alpha}{2} (\frac{\alpha}{2}-Q)$, with  $\alpha \in \mathbb{R}$. They are defined by:
\begin{equation}
\label{eq:vertex}
V_{\alpha}(x) := \lim_{\epsilon\to 0} \epsilon^{-\frac{\alpha^2}{2}}e^{\im \alpha (X_{g_0,\epsilon}(x)+c)},\quad x \in \D.
\end{equation} 
The simplest non-trivial correlation function of the theory is the $1$-point function:
\begin{align}
\label{def_1pt}
\langle V_{\alpha}\rangle=\int \;V_{\alpha}(0) \;e^{-\mathcal{S}^{\text{im}}_{\rm L}(\D,g_0,\phi)} \;\mathcal{D} \phi,
\end{align}
where $\int \mathcal{D} \phi $ denotes the path integral over $\phi$. The probabilistic construction of the above correlation function involves the imaginary Gaussian multiplicative chaos (GMC) on the circle $\mathbb{S}$:
\begin{equation}
\label{MGCC}
M_{\beta}^{\mathbb{S}} :=\lim_{\epsilon\to 0}\frac{1}{2\pi}\int_0^{2\pi} \big(  \; \epsilon^{-\frac{\beta^2}{8}} e^{\im \frac{\beta}{2} X_{g_0,\epsilon}(e^{\im \theta})} \big) d\theta,
\end{equation}
and the corresponding moment generating function:
\begin{equation}\label{eq:momgen}
\mathcal{G}^{\mathbb{S}}_{\beta}\left(\mu\right) =\mathbb{E}\big[e^{-\mu M_{\beta}^{\mathbb{S}}}\big].
\end{equation}
We restrict ourselves to the range of values $\beta \in (0,\sqrt{2})$. In that case, the limit in (\ref{MGCC}) is non-trivial and defines a random variable with finite $L^2$-moments, $\mathbb{E}[|M_{\beta}^{\mathbb{S}} |^2]<\infty$.
Outside of this range, the limit cannot be given sense as a random variable. The full phase diagram of the imaginary GMC was studied in \cite{lacoin2015}. 
Furthermore, for $\beta \in (0,\sqrt{2})$ it has been proven in  \cite{guillarmou2023} that the 
 random variable has exponential moments so that the expectation \eqref{eq:momgen} is well-defined.

\begin{figure}
\begin{tikzpicture}
  \begin{axis}[
    axis x line=center,
    axis y line=center,
    xlabel={$\text{Re}(c)$},
    ylabel={$\text{Im}(c)$},
    xmin=-30,
    xmax=70,
    ymin=-21,
    ymax=20,
    xtick={-21,21,63},
    ytick=\empty,
    xticklabels={$-\frac{2 \pi}{\beta}$,$\frac{2 \pi}{\beta}$,$\frac{6\pi}{\beta}$},
  legend style={
        at={(0.5,0.8)},
        anchor=west,
        font=\scriptsize
    }
  ]
    
    \addplot[name path=contour1,blue,thick]  table {reducedcontours1.dat};
    \addplot[name path=contour5l,blue!90,thick]  table {reducedcontours5_left.dat};    
    \addplot[name path=contour10l,blue!40,thick]  table {reducedcontours10_left.dat};    
    \addplot[name path=contour2,blue,thick] table {reducedcontours2.dat};
    \addplot[name path=contour5r,blue!90,thick]  table {reducedcontours5_right.dat};    
    \addplot[name path=contour10r,blue!40,thick]  table {reducedcontours10_right.dat};    
    \addplot[name path=floor, draw=none] coordinates {(-30,-25) (70,-25)};
    
    \addplot[color=blue!55, opacity=0.5] fill between[of=contour1 and contour5l];
    \addplot[color=blue!50, opacity=0.5] fill between[of=contour5l and contour10l];
    \addplot[color=blue!10, opacity=0.5] fill between[of=contour10l and floor];
\addplot[color=blue!55, opacity=0.5] fill between[of=contour2 and contour5r];
   \addplot[color=blue!50, opacity=0.5] fill between[of=contour5r and contour10r]; 
    \addplot[color=blue!10, opacity=0.5] fill between[of=contour10r and floor];

 \addplot[draw=none, fill=blue!55, opacity=0.5] coordinates {(0,0)};
\addlegendentry{$-5<\text{Re}[\mathcal{A}(c)]<0 $}
    \addplot[draw=none, fill=blue!50, opacity=0.5] coordinates {(0,0)};
\addlegendentry{$-10<\text{Re}[\mathcal{A}(c)]<-5$}
    \addplot[draw=none, fill=blue!10, opacity=0.5] coordinates {(0,0)};
\addlegendentry{$\text{Re}[\mathcal{A}(c)]<-10 $}

    \addplot[
      black,
      thick,
      postaction={
        decorate,
        decoration={
          markings,
          mark=at position 0.15 with {\arrow{>}},
          mark=at position 0.5 with {\arrow{>}},
          mark=at position 0.85 with {\arrow{>}}
        }
      }
    ] coordinates {(0,-25) (0,0) (42,0) (42,-25)};
    
    \node at (axis cs:21,2) {$\mathcal{U}$};

  \end{axis}
\end{tikzpicture}

\caption{In the $c$ complex plane, the regions where $\text{Re}\left[\mathcal{A}(c)\right]$ takes smaller and smaller negative values, where $\mathcal{A}(c)= \im \beta s c / 2 + \log{\mathcal{G}^{\mathbb{S}}_{\beta}(\mu e^{\im \beta/2 c})}$, are shaded in different blue tones. One can see that on the $\mathcal{U}$ contour (shown in black) the one-point function (\ref{1-pt_path_integral}) is well defined. The parameters used in the above picture are $\beta=0.6$, $s=0.37$ and $\mu=1$.}
\label{contours}
\end{figure}
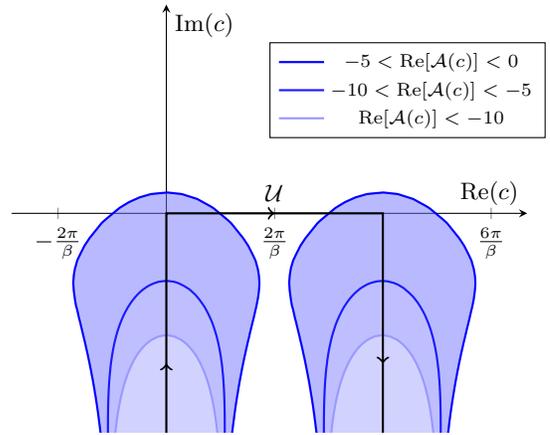

Splitting the integration $\int \mathcal{D} \phi \cdots = \int\;{\rm d}c \;\mathbb{E}[\dots]$, and using the definition (\ref{imagLFT}), the $1$-point correlation function (\ref{def_1pt}) can be expressed as:
\begin{align}
&\langle V_{\alpha}\rangle =\int_{\mathcal{C}}  \;e^{\im \left(\alpha-Q\right)c} \mathcal{G}^{\mathbb{S}}_{\beta}\big(\mu_B e^{\im \frac{\beta}{2} c}\big){\rm d}c.
\label{1-pt_path_integral}
\end{align}
The crucial point here is the choice of the contour $\mathcal{C}$ in the $c$ complex plane. As proposed in \cite{Harlow_2011}, the definition of the LFT$_{\im \beta}$ path integral could, in principle, be understood by the use of Morse theory. This would involve deforming the integration path into a sum of Lefschetz thimbles, which are the infinite-dimensional generalization of steepest descent contours \cite{witten2010}. Building from this idea, the semi-classical limit of the imaginary DOZZ function was studied and found to exhibit an analytic behavior similar to that of the inverse Gamma function \cite{Harlow_2011}. Note that, in the minisuperspace approximation where only the constant mode $c$ is integrated over, $M^{\mathbb{S}}_{\beta}=1$ and $\mathcal{G}^{\mathbb{S}}_{\beta}(\mu_B e^{\im \frac{\beta}{2} c})\to \exp(\mu_B e^{\im \frac{\beta}{2} c})$, the integrand in Eq.~(\ref{1-pt_path_integral}) is the same as the Gamma function. Taking $\mathcal{C}$ as a U-shaped (Hankel type) contour provides the integral representation of the inverse Gamma function. It was showed in \cite{Cao_2023} that U-shaped integrals are key to performing the analytic continuation of the lattice LFT$_{b}$ when $b$ approaches $\im \beta$. Inspired by these results, we set $\mathcal{C}=\mathcal{U}$, where $\mathcal{U}$ consists in the lines $[0,-\im \infty)$, $[0,\tfrac{4\pi}{\beta}]$ and $[\tfrac{4\pi}{\beta},\tfrac{4\pi}{\beta}-\im\infty)$, see Fig.\ref{contours}. It turns out that this choice of contour is perfectly consistent with the behavior of the Laplace transform $\mathcal{G}^{\mathbb{S}}_{\beta}\left(\mu\right)$ as we show rigorously that
\begin{equation}
\mathcal{G}^{\mathbb{S}}_{\beta}\left(\mu\right)\stackrel{\mu\to +\infty}{\sim}
\Gamma\big(1+\tfrac{\beta^2}{4}\big)^{\frac{4}{\beta^2}}\Gamma\big(\tfrac{4}{\beta^2}+1\big)\mu^{-\frac{4}{\beta^2}}.
\label{asy_G}
\end{equation}
A short proof of Eq.~(\ref{asy_G}) is provided in appendix. By taking $\mathcal{C}=\mathcal{U}$, Eq.~\eqref{1-pt_path_integral} then perfectly makes sense because the integral converges on the contour $\mathcal{U}$. It yields a non trivial quantity for $s=-\frac{2}{\beta}(\alpha-Q)$:
\begin{equation}   \label{circle_res1}
\begin{split}
  \langle V_{\alpha}\rangle^{\rm LFT_{   \im\beta }}=& \int_{\mathcal{U}} \;e^{\im \left(\alpha-Q\right)c} \mathcal{G}_\beta^{\mathbb{S}}\big(\mu_B e^{\im \frac{\beta}{2} c}\big){\rm d} c \\
= & \frac{4\pi }{\beta} \frac{e^{-\im \pi s } \mu_B^{s}}{\Gamma(1+s)} \frac{\Gamma(1+ s\frac{\beta^2}{4})}{\Gamma(1+\frac{\beta^2}{4})^s}.
\end{split}
\end{equation}
If $s\in \mathbb{N}$, the two vertical branches of the contour cancel each other and we are left with the horizontal contribution of the contour. This expression is computable with straightforward Fourier
analysis and yields:
\begin{align}
\label{circle_CG}
&\langle V_{\alpha}\rangle^{\rm LFT_{   \im\beta }}=\int_{0}^{\frac{4\pi}{\beta}}e^{-\im \frac{\beta}{2} s c} \; \mathbb{E}\big[e^{-\mu_B e^{\im \frac{\beta}{2} c }M_{\beta}^{\mathbb{S}}}\big] \, {\rm d}c\nonumber \\&=\frac{4\pi}{\beta}\frac{\left(-\mu_{B}\right)^{s}}{s!}\mathbb{E}\left[(M_{\beta}^{\mathbb{S}})^{s}\right]=\frac{4 \pi}{\beta}\frac{\left(-\mu_{B}\right)^{s}}{s!}\frac{\Gamma(1+ s\frac{\beta^2}{4})}{\Gamma(1+\frac{\beta^2}{4})^s}.
\end{align}
The expression \eqref{circle_CG} can be seen as the circular version of the Coulomb Gas integral (or CILT) and it is an evident extension of (\ref{circle_res1}) for real values of $s$. 
Moreover, we also observe below that, up to some normalization-dependent multipliers, the $\langle V_{\alpha}(0)\rangle^{\rm LFT_{   \im\beta }}$ satisfies the same shift relations as those satisfied by the real LFT$_{b }$ on the circle. This is not surprising: the correlation function of LFT theories are known to obey shift relations stemming from the insertion of degenerates states, which are assumed to be present in the theory. The behavior of these states depends  only on the structure of the Virasoro algebra and is therefore expected to remain invariant under the analytic continuation of the parameters $b$ and $\alpha$. The one-point function  of the real LFT$_{b }$, $\langle V_\alpha(0)\rangle^{\rm LFT_{b }}$, can be computed as the solutions of these shift relations  or via the standard probabilistic expression:
\begin{equation}\label{standard1pt}
\langle V_\alpha\rangle^{\rm LFT_{b }}=\frac{2}{b}\mu^{-\tilde{s}}\Gamma(1-\tfrac{b^2}{4})^{\tilde{s}}\Gamma(\tilde{s})\Gamma(1+\tilde{s}\tfrac{b^2}{4})
\end{equation}
with $\tilde{s}=\tfrac{2}{b}(\alpha-Q)$.
There is a simple relation between the $1$-point function of the standard theory $\langle V_\alpha\rangle^{\rm LFT_{b }}$, which can be formally continued to $b=\im\beta$, and the $1$-point function of our theory:
\begin{equation}
\label{2_LFTs}
 \langle V_{\im\alpha}\rangle^{\rm LFT_{b\to \im\beta }}=\frac{\im e^{\im \pi s}}{2\;\sin{\pi s}} \langle V_{\alpha}\rangle^{\rm LFT_{   \im\beta }}\ .
\end{equation}
This last expression echoes the relation between the real DOZZ formula and the imaginary DOZZ formula~\cite{Zam2005}, the coefficient $\frac{\im e^{\im \pi s}}{2\;\sin{\pi s}}$ playing the role of the step function in~\cite{RiSa14}. We observe also that $\langle V_{\im\alpha}\rangle^{\rm LFT_{b\to  \im\beta }}$ can be obtained directly from   \eqref{1-pt_path_integral} if one chooses as a contour $\mathcal{C}=\im \mathbb{R}$: by standard contour deformation, one has $(1-e^{2\im \pi s})\int_{\im \mathbb{R}}dc \cdots=\int_{\mathcal{U}} dc \cdots$, which yields the relation (\ref{2_LFTs}). Notice that the $\langle V_{\im\alpha}\rangle^{\rm LFT_{b\to  \im\beta }}$ for $\beta=\sqrt{2}$, i.e. $c_{\text{charge}}=1$,  was also derived in \cite{strominger2003}, while a study of the solutions of the shift relation for $c_{\text{charge}}\leq 1$ and for non vanishing bulk cosmological constant,  was considered recently in \cite{teresa22}. 

The $1$-point function of the imaginary LFT reveals that, depending on the choice of contour for integrating the constant mode, one obtains two strictly related theories. The first one corresponds to the analytic continuation of the real LFT$_{b}$ under $b\to \im \beta$. A characteristic feature of this solution is that it includes the Coulomb gas (CG) integrals as its residues when $s\in\mathbb{N}$. The second, the LFT$_{\im \beta}$, is defined using the contour $\mathcal{U}$ and constitutes a direct extension of the $s$-dimensional CG integrals to $s\in \mathbb{R}$. We have also analyzed and exactly solved a special bulk-boundary $2$-point function. In particular, we checked that thanks to the asymptotic behavior (\ref{bbasy}) the integral over the $\mathcal{U}$ contour is well-defined in this case as well (see supplementary material).

The results discussed so far pertain to the simplest correlation function and rely on the behavior of the imaginary chaos on the circle. 
To gain a deeper understanding,  it is necessary to investigate more complex correlation functions that depend on the imaginary GMC on a two-dimensional manifold. Among these, the $3$-point correlation function on the sphere $\mathbb{S}^2$ is the simplest and most significant. As mentioned in the introduction, it is well known that, unlike the $1$-point function, the $3$-point function of the LFT$_{b}$ does not admit an analytic continuation to $b\to \im \beta$ when $\beta^2\notin \mathbb{Q}$. However, we show that it is possible to define a path integral formulation of LFT$_{\im \beta}$ which extends the CG results and agrees with the alternative bootstrap solution defined for $c_{\text{charge}}\leq 1$, namely, the imaginary DOZZ structure constant.

We identify  $\mathbb{S}^2$  with the extended complex plane $\hat{\mathbb{C}}=\mathbb{C}\cup \{\infty\}$ that we equip with the metric $g_0= |z|_{+}^{-4} |d z|^2$, where $|z|_{+}=\text{max}(|z|,1)$. The volume form ${\rm dv}$ and the scalar curvature $K_g$ of $\hat{\mathbb{C}}_{g_{0}}$ are respectively given by ${\rm dv} = |z|^{-4}_{+}{\rm d}x{\rm d}y$ for $z=x+\im y$ and $K_g = 4 \delta_{|z|=1}$. 
We split $\phi(x)$ as  $\phi(x)= X_{g_0}(x) + c$, where $c$ is a complex constant and $X_{g_0}(x)$ is a real centered GFF on $(\hat{\mathbb{C}},g_0)$ with covariance $\mathbb{E}\left[X_{g_0}(x)X_{g_0}(y)\right]=-\log{|x-y|}+\log{|x|_{+}|y|_{+}} $. The vertex operator $V_{\alpha}(x)$ is still defined as in (\ref{eq:vertex}), where one replaces  the disc GFF by $X_{g_0}$.
The $n$-point correlation function is expressed as:
\begin{align}
&\Big\langle \prod_{i=1}^n V_{\alpha_i}(x_i)\Big\rangle=\int \; \prod_{i} V_{\alpha_i}(x_i)e^{-\mathcal{S}^{\text{im}}_{\rm L}(\hat{\mathbb{C}},g_0,\phi)} \mathcal{D}\phi .
\label{path_integral_sphere}
\end{align}
In this Letter, we restrict our focus to the three-point functions $C_{\boldsymbol{\alpha}}=\left< V_{\alpha_1}(0)V_{\alpha_2}(1)V_{\alpha_3}(\infty)\right>$, $\boldsymbol{\alpha}=\{\alpha_i\}$, which are the structure constants of the theory. The main object entering the definition of the functions $C_{\boldsymbol{\alpha}}$ is the imaginary chaos with insertion of charges:
\begin{equation}
M^{\hat{\mathbb{C}}}_{\beta,\boldsymbol{\alpha}} :=\frac{1}{2\pi}\lim_{\epsilon\to 0}\int_{\hat{\mathbb{C}}}  \; \frac{|z|^{\beta \alpha_1}|z-1|^{\beta \alpha_2}}{|z|^{\beta\bar{\alpha}}_{+}}\epsilon^{-\frac{\beta^2}{2}} e^{\im \beta X_{g_0,\epsilon}(x)} {\rm dv}, 
\end{equation}
where $\bar{\alpha}=\alpha_1+\alpha_2+\alpha_3$ and   $X_{g_0,\epsilon}$ is an ultraviolet regularization of the GFF at scale $\epsilon$. 

\begin{figure}
	\includegraphics[width=.99\linewidth]{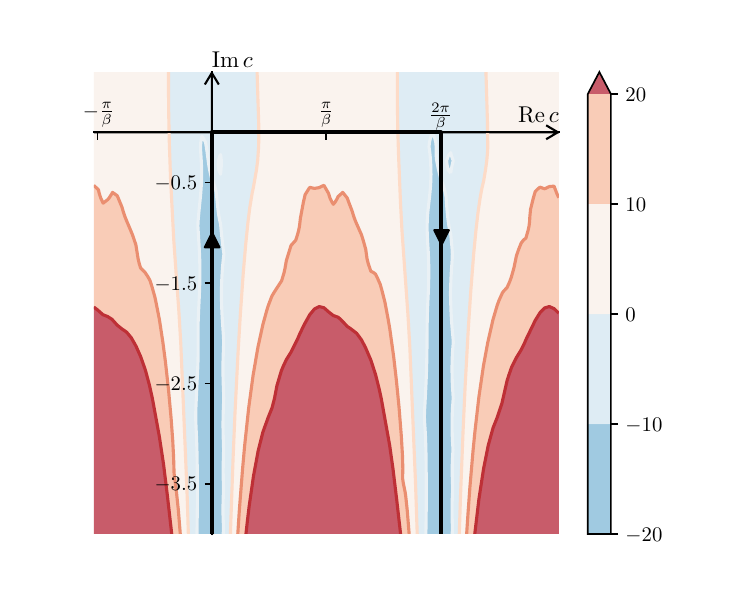}
	\caption{ Contour lines of the real part of $\im (\bar{\alpha}-Q) c+\log(\mathcal{G}^{\hat{\mathbb{C}}}_{\beta,\boldsymbol{\alpha}}(e^{\im \beta c}))$, as from numerical simulations with $N=12000$ independent samples of the chaos, $\beta=0.4$, $\alpha_1=\alpha_2=-4.0$, $\alpha_3=-2.4$. \label{contours2d}}
\end{figure}

The imaginary chaos $M^{\hat{\mathbb{C}}}_{\beta,\boldsymbol{\alpha}} $ is well defined and non trivial~\cite{lacoin2015} when $
\alpha_i> Q, \quad \beta^2< 2$.
Both conditions are needed to make the random variable probabilistically well defined, with finite $L^2$ moments \cite{guillarmou2023}.
Again, these conditions also ensure that $M^{\hat{\mathbb{C}}}_{\beta,\boldsymbol{\alpha}} $ has finite exponential moments in such a way that the corresponding generating function 
\begin{equation}
\mathcal{G}^{\hat{\mathbb{C}}}_{\beta,\boldsymbol{\alpha}}(\mu)=\mathbb{E}\big[e^{-\mu M^{\hat{\mathbb{C}}}_{\beta,\boldsymbol{\alpha}}}\big]
\end{equation} 
is well defined~\cite{guillarmou2023}. Analogously to the circle case, we define the LFT$_{\im \beta}$ by integrating the constant mode $c$ over an U-type contour,  $\mathcal{U}$, consisting in the lines $[0,-\im \infty)$, $[0,\tfrac{2\pi}{\beta}]$ and $[\tfrac{2\pi}{\beta},\tfrac{2\pi}{\beta}-\im\infty)$ -- see Fig.~\ref{contours2d}. The structure constants of the theory are:
\begin{align}
\label{defC}
C_{\boldsymbol{\alpha}}=\int_{\mathcal{U}} \;e^{\im \left(\bar{\alpha}-2 Q\right)c} \;  \mathcal{G}^{\hat{\mathbb{C}}}_{\beta,\boldsymbol{\alpha}}(\mu e^{\im \beta c}) {\rm d}c \ .
\end{align}
We conjecture that the $\mathcal{G}^{\hat{\mathbb{C}}}_{\beta,\boldsymbol{\alpha}}(\mu)$ behaves similarly to the circle case $\mathcal{G}^{\mathbb{S}}_{\beta}(\mu)$. We computed numerically the function $\mathcal{G}^{\hat{\mathbb{C}}}_{\beta,\boldsymbol{\alpha}}(\mu)$, for different values of $\beta$ and the $\alpha_{i}$'s, satisfying the conditions above.

In Fig.~\ref{contours2d}, we show the contour lines of $\im (\bar{\alpha}-Q) c+\log \mathcal{G}^{\hat{\mathbb{C}}}_{\beta,\boldsymbol{\alpha}}(e^{\im \beta c})$, supporting our conjecture that the definition (\ref{defC}) perfectly makes sense.  

Let us rewrite Eq.~\eqref{defC} in a more suggestive form:
\begin{align}
C_{\boldsymbol{\alpha}}  =&\left(-1+e^{-2 \im \pi s}\right)\int_{0}^{-\im \infty}\; \;e^{-\im \beta sc} \; \mathcal{G}^{\hat{\mathbb{C}}}_{\beta,\boldsymbol{\alpha}}\left(\mu e^{\im \beta c}\right){\rm d}c+\nonumber \\
&+\int_{0}^{\frac{2\pi}{\beta}}\; \;e^{-\im \beta sc}\mathcal{G}^{\hat{\mathbb{C}}}_{\beta, \boldsymbol{\alpha}}\left(\mu e^{\im \beta c}\right){\rm d}c
\label{def_struct}
\end{align}
with $s=\frac{2 Q-\bar{\alpha}}{\beta}$.
When $s\in \mathbb{N}$, the definition \eqref{def_struct} includes as a special case the diagonal sector of the compactified  imaginary Liouville (CILT) for which the structure constants are given by the Dotsenko-Fateev integrals given below (\ref{cg}).
The LFT$_{\im \beta}$, as defined above, represents an extension of the Coulomb gas correlation functions -- and hence of the CILT -- to real values of the parameter $s$. Fortunately, we can compare numerically the $3$-point function (\ref{def_struct}) to a solution for the $3$-point function obtained via the bootstrap approach: the imaginary DOZZ structure constant  $C^{\text{ImDOZZ}}_{\boldsymbol{\alpha}}$. This constant was introduced in \cite{Schomerus_2003,kp05a,Zam2005,Dotsenko_2016} as the unique solution to a pair of shift equations at $c_{\text{charge}}\leq 1$. Moreover, it can be derived as the unique analytic continuation of the Dotsenko-Fateev integrals with $3$ electric charges~\cite{kp05a,psvd13,Dotsenko_2016} consistent with the duality $\beta\to -\frac{4}{\beta}$. This is precisely why our theory must coincide with this bootstrap solution:
\begin{equation}
\label{testtheory}
C^{\phantom{\text{I}}}_{\boldsymbol{\alpha}}=C^{\text{ImDOZZ}}_{\boldsymbol{\alpha}} .
\end{equation}
Our numerical simulations exhibit excellent agreement with Eq.~\eqref{testtheory}. The quality of that agreement is illustrated in Fig.~\ref{fig:match_overview}. Further details about numerical simulations are given in the supplementary material below, including a discussion of their range of validity.

\begin{figure}
	\includegraphics[width=.99\linewidth]{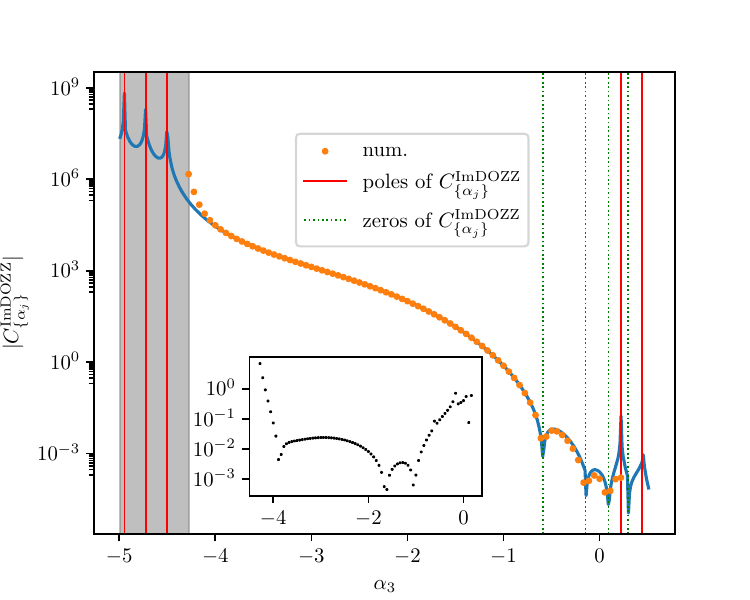}
	\caption{\label{fig:match_overview} Overview of the match between the imaginary DOZZ formula and the numerical simulations of LFT$_{\im\beta}$ three-point function. The relative deviation is shown in the insert. The area $\alpha_3\leq Q$ is shaded in gray, and the poles and zeros of $C^\text{ImDOZZ}$ are marked by vertical lines. See the supplementary material for further details.}
\end{figure}

For a general surface $(\Sigma,g)$, the same heuristics lead us to propose the following definition for the correlation functions (with $s=\sum_{j=1}^n\alpha_j -Q\chi(\Sigma)$):
\[
\begin{gathered} 
\Big\langle \prod_{j=1}^nV_{\alpha_j}(x_j)\Big\rangle:=F_{g,{\bf x},\boldsymbol{\alpha}}\int_{\mathcal{C}} e^{\im sc} \, \mathbb{E}[e^{\mu e^{\im \beta c} 
M_{\beta,{\bf x},\boldsymbol{\alpha}}^{\Sigma,g}}]\;{\rm d}c \ ,\\  
M_{\beta,{\bf x},\boldsymbol{\alpha}}^{\Sigma,g}:=\lim_{\epsilon\to 0}\int_\Sigma  e^{-\beta Q u(x)}
\epsilon^{-\frac{\beta^2}{2}}e^{\im \beta X_{g,\epsilon}(x)}\;{\rm dv}(x)\end{gathered}\]
with $\mathcal{C}$ the same U-shaped contour as before, $X_{g,\epsilon}$ the $\epsilon$-regularization of the GFF $X_g$ on $(\Sigma,g)$,
$F_{g,{\bf x},\boldsymbol{\alpha}}$ some explicit term  and $u$ a function with local singularities $\prod_{j=1}^n|x-x_j|^{\beta \alpha_j}$ at $x_j$, see below. We postulate that $\mathbb{E}[e^{-\mu
M_{\beta,{\bf x},\boldsymbol{\alpha}}^{\Sigma,g}}]\sim \mu^{-\lambda}$ as $\mu\to +\infty$ for some $\lambda>0$ depending on $\beta$ and the $\alpha_i$'s, which implies that the integral converges under the Seiberg type bounds 
\[ \alpha_j >Q ,\quad s=\sum_{j=1}^n\alpha_j -Q\chi(\Sigma)<\beta \lambda,\]

In summary, we have proposed a definition of the imaginary LFT$_{\im \beta}$ correlation functions via a path integral in which the zero mode is integrated along a Hankel-type contour. In the case of the $3$-point function, this leads to the expression (\ref{defC}). The validity of our construction relies on the asymptotic behavior of the  moment generating function of the  imaginary GMC, which we confirmed in the case of imaginary chaos on the circle, where exact results are available. Furthermore, we did test our theory through a highly non-trivial comparison between numerical simulations and the imaginary DOZZ structure constant. Our results not only provide a direct probabilistic definition of the imaginary DOZZ structure constant, but also highlight the existence of a conformal field theory that arises as the analytic continuation of CILT. An important open question raised by our results concerns the factorization properties of the $4$-point function. Notably, the only known non-compact bootstrap solution featuring $C^{\text{ImDOZZ}}$ as structure constants, introduced in \cite{RiSa14}, is not an extension of the Coulomb gas integrals. We leave this question for future investigation.

\clearpage
\appendix
\onecolumngrid
\section{Imaginary GMC on the circle: exact results}
\label{asyG}
We give a derivation of Eq.~\eqref{asy_G}. We use the following property of the GFF under consideration:
\begin{equation}
\lim_{\epsilon\to 0}\;\epsilon^{-\frac{\beta^2}{4}}\mathbb{E}\big[e^{\im\frac{\beta}{2} X_{g_0,\epsilon}(e^{\im \theta_1})} e^{\im\frac{\beta}{2} X_{g_0,\epsilon}(e^{\im \theta_2})}\big]=\left|e^{\im \theta_1}-e^{\im \theta_2}\right|^{\frac{\beta^2}{2}}.
\end{equation}
An explicit implementation of the UV regularization $\epsilon$ is provided in supplementary material below. By Wick's theorem, the $n$-th moment of the imaginary chaos $\mathbb{E}[(M^\mathbb{S}_\beta)^{n}]$, $n\in \mathbb{N}$ is expressed in terms of the integral:
\begin{equation}
\mathbb{E}\left[(M^\mathbb{S}_\beta)^n\right] = \int_{ [0,2\pi]^n }  \prod_{1 \leq j<k \leq n } \left|e^{\im \theta_j}-e^{\im \theta_k}\right|^{ \frac { \beta^2 }{ 2 } }\prod_{i=1}^n \frac{ {\rm d}\theta_ i}{2\pi}.
\end{equation}
This is a Selberg-type integral whose value is given by \cite{reviewselberg}:
\begin{equation}
\mathbb{E}\left[(M^\mathbb{S}_\beta)^n\right] =\Gamma\big( 1 + \tfrac{ n\beta ^2 }{ 4 }\big) \Gamma\big( 1 + \tfrac{ \beta ^2 }{ 4 } \big)^{-n}.
\end{equation}
This allows to give an integral representation of the moment-generating function $\mathcal{G}^{\mathbb{S}}_{\beta}\left(\mu\right)$:
\begin{equation}\label{eq:laplace_integral-appendix}
\mathcal{G}^{\mathbb{S}}_{\beta}\left(\mu\right) =\sum_{n=0}^{\infty} \frac{ (-\mu )^n } { n! \; \Gamma(1 + \frac{ \beta ^2 }{ 4 })^n  } \Gamma(1 + \tfrac{ n\beta ^2 }{ 4 })=\sum_{n=0}^{\infty} \frac{ (-\mu) ^n } { n! \; \Gamma(1 + \frac{ \beta ^2 }{ 4 })^n  }\int _ 0 ^ { + \infty }  t^{\frac{n\beta^2}{4}} e^{-t} {\rm d}t= \int _ 0 ^{ + \infty } e^{ -\overline{\mu} t ^{\frac{ \beta ^2 }{4}} - t} {\rm d}t
\end{equation}
for $\overline{\mu}=\mu/\Gamma(1+\frac{\beta ^2}{4})$, where we use the integral expression of the Gamma function. 

In the limit $\mu \to -\infty$,  the above integral is dominated by its saddle-point: $\mathcal{G}^{\mathbb{S}}_{\beta}$  grows  exponentially in this $\mu$-direction, see the supplementary material. In the limit $\mu\to \infty$, the integral has no real saddle-point, but a change of variable produces:
\begin{equation}
\mathcal{G}^{\mathbb{S}}_{\beta}\left(\mu\right)= \overline{\mu}^{-\frac{4}{\beta^2}}\int_0^\infty e^{-t^{\frac{\beta^2}{4}}-\overline{\mu}^{-\frac{4}{\beta^2}}t}{\rm d}t
\end{equation}
and expanding the term $\exp(-\overline{\mu}^{-\frac{4}{\beta^2}}t)$ in power series leads to the asymptotics \eqref{asy_G}. 
We have also solved the case where one inserts in the correlation function, besides the vertex $V_{\alpha}(0)$ at the origin, a vertex $V_{\eta}(1)$ at the boundary, with $\eta=\beta/2$. In the bulk-boundary correlation function the main role is played by 
\begin{equation}
\mathcal{G}^{\mathbb{S}}_{\beta,\eta}(\mu)= \mathbb{E}\big[e^{-\mu M_{\beta,\eta}^{\mathbb{S}}}\big],\quad \text{where}\;M_{\beta,\eta}^{\mathbb{S}} =\frac{1}{2\pi}\int_0^{2\pi} |e^{\im \theta}-1|^{\beta \eta} e^{\im \frac{\beta}{2} X_{g_0}(e^{\im \theta})} d\theta \ .
\end{equation} 
Setting $\eta=\beta/2$, we have when $\mu\to\infty$:
\begin{equation}
\label{bbasy} 
\mathcal{G}^{\mathbb{S}}_{\beta,\eta}(\mu) \sim \frac{4}{\beta^2}\Gamma(\frac{4}{\beta^2}+1)\Gamma(1+\frac{\beta^2}{4})^{\frac{4}{\beta^2}}\mu^{-\frac{4}{\beta^2}-1} \ .
\end{equation}

\section{Imaginary DOZZ structure constant}
The $C^{\text{ImDOZZ}}_{\boldsymbol{\alpha}}$  is expressed in terms of functions, $\Upsilon_{\frac{\beta}{2}}(x)$, which are strictly related to the double Gamma function, see Sec. 7 of~\cite{guillarmou2023}. Using the same field normalization as in~\cite{guillarmou2023},  $C^{\text{ImDOZZ}}_{\{\alpha_i\}}$ is:
\begin{align}
     C^{\text{ImDOZZ}}_{\boldsymbol{\alpha}} =&
\frac{2\pi}{\beta}\Big(\frac{-\mu}{\gamma(\frac{\beta^2}{4})}\Big)^{s}
\Big(\frac{\beta}{2}\Big)^{Q \beta s} \frac{\Upsilon_{\frac{\beta}{2}}(Q
+\frac{\beta}{2}-\frac{\overline{\alpha}}{2})}{\Upsilon_{\frac{\beta}{2}}(\frac{\beta}{2})} \frac{
\prod_{i=1}^3\Upsilon_{\frac{\beta}{2}}(\frac{\beta}{2}-\frac{\overline{\alpha}}{2}+\alpha_i)
     }{
\prod_{i=1}^{3}\Upsilon_{\frac{\beta}{2}}(\frac{2}{\beta}+\alpha_i)
     },
     \label{eq:Im_DOZZ}
\end{align}
where $\gamma(x)=\Gamma(x)/\Gamma(1-x)$.
Using the recursion properties of the $\Upsilon_{\frac{\beta}{2}}(z)$ functions:
\begin{align}
&\Upsilon_{\frac{\beta}{2}}\big(z+\frac{\beta}{2}\big)=\gamma\big(\frac{\beta}{2} z\big)\big(\frac{\beta}{2}\big)^{1-\beta z}\;\Upsilon_{\frac{\beta}{2}}\left(x\right),\quad \Upsilon_{\frac{\beta}{2}}\big(z+\frac{2}{\beta}\big)=\gamma\big(\frac{2}{\beta} z\big)\big(\frac{\beta}{2}\big)^{\frac{4}{\beta}z-1}\Upsilon_{\frac{\beta}{2}}(x),
\end{align}
one can show that when $s\in\mathbb{N}$, Eq.~\eqref{eq:Im_DOZZ} yields the value of the Coulomb gas integral with three electric charges:
\begin{equation}\label{cg}
\begin{split}
C^\text{CG}_{\boldsymbol{\alpha}} \stackrel{s\in \mathbb{N}}=& \int_{0}^{\frac{2\pi}{\beta}}\; {\rm d}c\;e^{-\im \beta s}\mathcal{G}^{\hat{\mathbb{C}}}_{\beta, \boldsymbol{\alpha}}
=\frac{2\pi}{\beta}\frac{\left(-\mu\right)^s}{s!}\mathbb{E}[(M^{\hat{\mathbb{C}}}_{\beta,\boldsymbol{\alpha}})^s]=\nonumber \\
\stackrel{s\in \mathbb{N}}=& \frac{2\pi}{\beta}\frac{\left(-\mu\right)^s}{s!} \int_{\mathbb{C}^s} \prod_{i=1}^{s} {\rm d}\xi_i \;\prod_{i=1}^{s}|\xi_i|^{\alpha_1 \beta} |\xi_i-1|^{\alpha_2 \beta}\prod_{i<j=1}^{s}|\xi_i-\xi_j|^{\beta^2}=\frac{2\pi}{\beta}\Big(\frac{-\mu}{\gamma(\frac{\beta^2}{4})}\Big)^s\frac{\prod_{j=1}^{s}\gamma(j \frac{\beta^2}{4})}{\prod_{j=1}^{s-1}\prod_{i=1}^{3} \gamma(-\frac{\beta \alpha_i}{2}-j\frac{\beta^2}{4})}
\end{split}
\end{equation}

\section*{Supplementary material}\label{supp}
We provide here additional details, results, and observations.

\section{ Liouville theory on the circle }
Many insights in the Lagrangian description of the imaginary Liouville theory, and its relation with the standard Liouville theory, can be gained by considering the simpler case where we ignore bulk contributions and set the spacetime manifold to be the unit circle. 

\subsection{Real Liouville theory on the circle }
Let us recall the Lagrangian description of the standard Liouville theory on a Riemann surface with boundary. Consider a compact oriented surface $\Sigma$ with boundary $\partial \Sigma$ and equipped with a Riemannian metric $g$. We denote by ${\rm dv}$ and  ${\rm d}\ell$ the volume forms on $\Sigma$ and $\partial \Sigma$, and by $K_g$ and $k_g$ the scalar curvature on $\Sigma$ and geodesic curvature on $\partial \Sigma$. The (real) Liouville theory  LFT$_{  \gamma}$ on $\Sigma$  is defined through the action
\begin{align}
	\label{realLFT}
	\mathcal{S}^{{\rm L}}_{\Sigma,g}(\phi)=&\frac{1}{4\pi}\int_{\Sigma} \left( |d \phi|^2_{g}+  Q K_g \phi  + \mu e^{  \gamma \phi}\right){\rm dv}+\frac{1}{2\pi}\int_{\partial \Sigma} \left(  Q k_g \phi+\mu_B e^{ \frac{\gamma}{2} \phi}\right)\;{\rm d} \ell,
\end{align}
where $\phi(x)$ is the fundamental field, and $\mu$ and $\mu_B$ are non-negative numbers known as the bulk and boundary cosmological constants. The parameter $\gamma\in(0,2)$ is related to the background charge $ Q:=\frac{\gamma}{2} + \frac{2}{\gamma}$, such that the corresponding quantum field theory enjoys conformal invariance with central charge $c_{{\rm L}}=1+6 Q^2 \geq 25$. Expectation values of functionals of the field, $\langle F(\phi)\rangle$, are formally defined in terms of the path integral: $\langle F(\phi)\rangle = \int F(\phi)e^{-\mathcal{S}^{\rm L}_{\Sigma,g}(\phi)}\mathcal{D} \phi $, where the integration ranges over a set of (generalized) functions $\phi:\Sigma\to\R$.

Following an active line of research in mathematics, a rigorous construction of the real Liouville theory with a boundary was proposed in~\cite{huang2017liouvillequantumgravityunit} (see~\cite{review24} for an introduction to the probabilistic construction of Liouville theory). The Liouville field makes sense as a random distribution: it is a massless free field or, in the mathematical language, a \textit{Gaussian free field} (GFF)~\cite{sheffield2006gaussianfreefieldsmathematicians}, shifted by a real additive constant $c$ called the zero mode.

The basic objects in the probabilistic construction of the  boundary LFT$_{\gamma}$ on the unit disc $\D$ with flat metric $g_0=|dz|^2$ involve an ultraviolet regularization  $X_{g_0,\epsilon}$ at scale $\epsilon$ of the Neumann  GFF $X_{g_0}$, a random (Gaussian) distribution with covariance $\mathbb{E} [ X_{g_0}(x)X_{g_0}(y)]  =-\log{|x-y||1-x\bar y|}$. The boundary potential then makes sense in terms of the Gaussian multiplicative chaos (GMC) on the circle $\mathbb{S}$:
\begin{equation}
	\label{GMCreal}
	M_{\gamma}^{\mathbb{S}} :=\lim_{\epsilon\to 0}\frac{1}{2\pi}\int_0^{2\pi} \; \epsilon^{\frac{\gamma^2}{8}} e^{  \frac{\gamma}{2} X_{g_0,\epsilon}(e^{\im \theta})}  {\rm d}\theta,
\end{equation}
where the limit is non trivial for $\gamma\in (0,2)$. In the case $\mu=0$, for real-valued bounded continuous functionals $F$ on distributions, the expectation value which plays the role of path integral is given by the probabilistic quantity
$$\langle F\rangle:=\int_\mathbb{R} \E\Big[e^{-Q c}F(c+X_{g_0})e^{-\mu_B e^{\frac{\gamma}{2}c}M_{\gamma}^{\mathbb{S}}}\Big]\;{\rm d}c.$$
The primary fields of the  LFT$_\gamma$ are the vertex fields $V_{\alpha}$ with scaling dimension $\Delta = \frac{\alpha}{2} (\tilde Q-\frac{\alpha}{2})$, with  $\alpha \in \mathbb{R}$. They are defined by:
\begin{equation}
	\label{eq:vertex-supp}
	V_{\alpha}(x):= \lim_{\epsilon\to 0} \epsilon^{\frac{\alpha^2}{2}}e^{  \alpha (X_{g_0,\epsilon}(x)+c)},\quad x \in \mathbb{D}.
\end{equation} 
To probe the connections between LFT$_\gamma$ and its imaginary counterpart, it is instructive to compute the simplest  non-trivial correlation function, namely the one-point function $\langle V_{\alpha}(0)\rangle$, whose probabilistic expression yields after zero mode integration for $\alpha> Q$:
\begin{equation}
	\langle V_\alpha(0)\rangle =\frac{2}{\gamma}\mu_B^{-\tfrac{2}{\gamma}(\alpha-Q)}\Gamma\big(\tfrac{2}{\gamma}(\alpha- Q)\big)\E\big[\left(M^\mathbb{S}_{\gamma}\right)^{-\tfrac{2}{\gamma}(\alpha- Q)}\big].
\end{equation}
A formula for the $1$-point correlation function is provided by \cite{remy_integrability_2022}, which was derived from a closed-form expression for the moments of the Gaussian multiplicative chaos (GMC) on the circle. That formula
\begin{equation}
	\E[(M_\gamma^{\mathbb{S}})^{-\tfrac{2}{\gamma}(\alpha-Q)}]=\Gamma\big(1+\tfrac{\gamma}{2}(\alpha-Q)\big)\Gamma\big(1-\tfrac{\gamma^2}{4}\big)^{\tfrac{2}{\gamma}(\alpha-Q)},
\end{equation}
conjectured by Fyodorov and Bouchaud~\cite{fyodorov_freezing_2008} and later proved by Rémy~\cite{remy_fyodorov-bouchaud_2020}, has a closely related counterpart for the one-dimensional imaginary GMC -- as we will see later. We get the two shift equations:
\begin{equation}
	\langle V_{\alpha+\frac{\gamma}{2}}(0)\rangle =  A( \tfrac{\gamma}{2},\alpha,\mu_B)\langle V_{\alpha}(0)\rangle ,\qquad \langle V_{\alpha+\frac{2}{\gamma}}(0)\rangle 
	=A(\tfrac{2}{\gamma},\alpha,\tilde\mu_B)
	\langle V_{\alpha}(0)\rangle 
\end{equation}
where the coefficient $A$ and the dual cosmological constant $\tilde \mu_B$ are given by 
$$A(\chi,\alpha,\mu):=\frac{1}{\mu \chi^2}\frac{\Gamma(1-\chi^2)\Gamma(\chi\alpha)}{\Gamma(\chi(\alpha-Q))},\qquad \tilde\mu_B:=\frac{\mu_B^{\frac{4}{\gamma^2}}\Gamma(1-\tfrac{4}{\gamma^2})}{   \Gamma(1-\tfrac{\gamma^2}{4}) ^{\frac{4}{\gamma^2}}}.$$
The terminology comes from the \textbf{duality} relation 
$$\langle V_{\alpha}(0)\rangle_{\gamma,\mu_B}= \langle V_{\alpha}(0)\rangle_{\frac{4}{\gamma},\tilde\mu_B}$$
where the subscripts mean that we replaced $\gamma \to \frac{4}{\gamma}$, and $\mu_B \to \tilde\mu_B$.

\subsection{Imaginary Liouville theory on the circle }

The action for the imaginary Liouville theory is the same up to a rotation of the parameters in the complex plane: for $\mu,\mu_B$ non-negative 
\begin{align}
	\label{imagLFT-supp}
	\mathcal{S}^{{\rm iL}}_{\Sigma,g}(\phi)=&\frac{1}{4\pi}\int_{\Sigma} \left( |d \phi|^2_{g}+\im Q K_g \phi  + \mu e^{\im \beta \phi}\right)\; {\rm dv} +\frac{1}{2\pi}\int_{\partial \Sigma} \left(\im Q k_g \phi+\mu_B e^{\im \frac{\beta}{2} \phi}\right){\rm d} \ell,
\end{align}
where, again, $\phi(x)$ is the fundamental field, $\mu,\mu_B \geq 0$ are  the bulk and the boundary cosmological constants (the other notations remain the same as in the real case). The parameter $\beta \in \R$ is a real parameter and this time one sets   $  Q=\frac{\beta}{2} - \frac{2}{\beta}$ to get   conformal invariance with central charge $c_{\text{iL}}=1-6 Q^2 \leq 1$. We again focus on the situation where the bulk cosmological constantis set to $\mu=0$.

Our probabilistic proposal for the path integral of the imaginary theory, denoted here by  LFT$_{\im \beta}$ for $\beta\in\mathbb{R}$, involves the imaginary GMC on the circle $\mathbb{S}$ ($X_{g_0}$ is still the Neumann GFF on $(\mathbb{D},g_0=|dz|^2)$)
\begin{equation}
	\label{MGCC-supp}
	M_{\beta}^{\mathbb{S}} :=\lim_{\epsilon\to 0}\frac{1}{2\pi}\int_0^{2\pi}   \; \epsilon^{-\frac{\beta^2}{8}} e^{\im \frac{\beta}{2} X_{g_0,\epsilon}(e^{\im \theta})}  {\rm d}\theta.
\end{equation}
Note that the limit in Eq.~\eqref{MGCC-supp} is non trivial in the range $\beta \in (0,\sqrt{2})$ \cite{lacoin2015} and defines a random variable. Its generating function is well-defined~\cite{guillarmou2023}:
\begin{equation}\label{eq:momgen-supp}
	\mathcal{G}^{\mathbb{S}}_{\beta}\left(\mu\right) = \mathbb{E}\big[e^{-\mu M_{\beta}^{\mathbb{S}}}\big],
\end{equation}
and will be instrumental in the sequel. In particular, it is essential to understand the behaviour of the generating function with respect to the complex parameter $\mu\in\mathbb{C}$. Restricting our analysis to the circle allows for exact computations of this function.

\subsection{ Laplace transform of the imaginary GMC on the circle }

First, we give an integral expression for the generating function $\mathcal{G}_\beta^{\mathbb{S}}$. Let $\beta \in (0,\sqrt{2})$ and $n\in\N$. The $n$-th moment of the imaginary chaos on the circle is
\begin{equation}
\E [(M^\mathbb{S}_\beta)^n] = \int_{ [0,2\pi]^n } \prod_{1 \leq j<k \leq n } \left|e^{i\theta_j}-e^{i\theta_k}\right|^{ \frac { \beta^2 }{ 2 } } \prod_{i=1}^n \frac{ d\theta_ i}{2\pi} = \frac{\Gamma\big( 1 + \frac{ n\beta ^2 }{ 4 } \big)}{\Gamma\big( 1 + \frac{ \beta ^2 }{ 4 } \big)^n} \ ,
\end{equation}
where we used successively Wick's theorem and the integral formula known as Selberg integral~\cite{selberg_bemerkninger_1944} or \cite[eq (1.17)]{reviewselberg}. Then, we make use of the standard integral expression of the Gamma function: for $z>0$, $\Gamma(z) = \int_0^{+\infty} t^{z-1}e^{-t}dt$. We find the following integral formula:
\begin{align}
\mathcal{G}^{\mathbb{S}}_{\beta}\left(\mu\right) &= \E [e^{ -\mu M^\mathbb{S}_\beta }] = \sum_{n=0}^{\infty} \frac{ (-\mu) ^ n }{ n! } \E [(M^\mathbb{S}_\beta)^n] = \sum_{n=0}^{\infty} \frac{ (-\mu) ^n } { n! \; \Gamma\big( 1 + \frac{ \beta ^2 }{ 4 } \big)^n  }  \int _ 0 ^ { + \infty }  t^{\frac{n\beta^2}{4}} e^{-t}{\rm d}t \nonumber \\
&= \int _ 0 ^{ + \infty } e^{ -\overline{\mu} t ^{\frac{ \beta ^2 }{4}} - t}{\rm d}t \label{eq:laplace_integral-supp}
\end{align}
where $\overline{\mu} = \mu/\Gamma\big( 1 + \frac{ \beta ^2 }{ 4 } \big)$. It is not difficult to derive its asymptotic behavior in the complex plane.  When $\mu$ goes to $+\infty$ with positive real part, $\mathcal{G}^{\mathbb{S}}_{\beta}(\mu)$ has a full asymptotic expansion in powers of $\mu^{-4/\beta^2}$:
\begin{align}
	\int _ 0 ^{ + \infty } e^{ -\overline{\mu} t ^{\frac{ \beta ^2 }{4}} - t} {\rm d}t &= \overline{\mu}^{-\frac{4}{\beta^2}}\int_0^\infty e^{-t^{\frac{\beta^2}{4}}-\overline{\mu}^{-\frac{4}{\beta^2}}t}dt=\sum_{k=0}^N \overline{\mu}^{-\frac{4(k+1)}{\beta^2}} \int_0^\infty\frac{e^{-t^{\frac{\beta^2}{4}}}(-t)^k}{k!}dt +\mathcal{O}( \overline{\mu}^{-\frac{4(N+2)}{\beta^2}})\nonumber\\
	&= \frac{4}{\beta^2}\sum_{k=0}^N\frac{(-1)^k}{k!}\Gamma(\tfrac{4(k+1)}{\beta^2}) \overline{\mu}^{-\frac{4(k+1)}{\beta^2}} +\mathcal{O}( \overline{\mu}^{-\frac{4(N+2)}{\beta^2}})\label{expansionG}
\end{align}
for all $N\in \N$. On the other hand, when $\mu\to -\infty$ along the real axis,
\begin{align}
	\int_0 ^{ + \infty } e^{ -\overline{\mu} t ^{\frac{ \beta ^2 }{4}} - t} {\rm d}t =& (-\overline{\mu})^{\frac{4}{4-\beta^2}} \int_0^{+\infty} e^{(-\overline{\mu})^{\frac{4}{4-\beta^2}} ( s^{\frac{\beta^2}{4}}-s)} {\rm d}s
	\label{eq:integral_saddle}
\end{align}
The integral~\eqref{eq:integral_saddle} is dominated by the saddle-point $s_\ast = ( \frac{\beta^2}{4})^{\frac{4}{4-\beta^2}}$. It has the asymptotic behavior:
\begin{equation}
	\mathcal{G}^{\mathbb{S}}_{\beta}\left(\mu\right) \underset{\mu\to-\infty}{=} C  (-\overline{\mu})^{\frac{4}{4-\beta^2}-\frac{1}{2}} e^{(-\frac{\overline{\mu}\beta^2}{4})^{\frac{4}{4-\beta^2}}} ( 1 + o(1))
\end{equation}
where $C$ is an unimportant constant. The Laplace transform with an additional $\beta/2$ insertion can be dealt with by the same argument, or by differentiating $\partial_{\mu}\langle  V_\alpha(0)\rangle $ : 
for $M^\mathbb{S}_{\beta,\beta/2}=\frac{1}{2\pi}\int_0^{2\pi}|1-e^{i\theta}|^{\beta/2}e^{i\frac{\beta}{2}X_{g_0}^{\D}(e^{i\theta})}{\rm d}\theta$
we find that 
\[\E [e^{ -\mu M^\mathbb{S}_{\beta,\beta/2}}] \sim\frac{4}{\beta^2\Gamma(1+\beta^2/4)}\sum_{k=0}^N\frac{(-1)^k}{k!}
\overline{\mu}^{-\frac{4(k+1)}{\beta^2}-1}\Gamma(\tfrac{4(k+1)}{\beta^2}+1)+\mathcal{O}( \overline{\mu}^{-\frac{4(N+2)}{\beta^2}-1}). \]

\subsection{The $1$-point correlation function} 
Using the definition (\ref{imagLFT-supp}) and splitting the integration $\int \mathcal{D} \phi \cdots = \int\;dc \;\mathbb{E}[\dots]$,   the one-point correlation function can be expressed as:
\[
\langle V_{\alpha}\rangle=\int_{\mathcal{C}} e^{\im \left(\alpha-Q\right)c} \mathcal{G}^{\mathbb{S}}_{\beta}\left(\mu_B e^{\im \frac{\beta}{2} c}\right)\;{\rm d}c
\]
where the choice of integration contour $\mathcal{C}$ for the zero mode turns out to be crucial, as pointed out in the main text. We choose the U-shaped Hankel type contour consisting  in the lines $[0,-\im \infty)$, $[0,\tfrac{4\pi}{\beta}]$ and $[\tfrac{4\pi}{\beta},\tfrac{4\pi}{\beta}-\im\infty)$. This integral is well defined for $\alpha>Q$ due to the asymptotics \eqref{expansionG} of the moment generating function.

One can perform an exact computation of this $1$-point function by contour deformation. Indeed, one can deform the contour in the expression~\eqref{1-pt_path_integral} to integrate instead over the contour $\mathcal{C}_R$, for $R>0$,  formed by the lines $[\im R,-\im \infty)$, $[\im R ,\tfrac{4\pi}{\beta}+\im R]$ and $[\tfrac{4\pi}{\beta}+\im R,\tfrac{4\pi}{\beta}-\im\infty)$. For $\alpha>Q$ and taking the limit $R\to\infty$, one gets
$$\left< V_{\alpha}(0)\right>=\im (1-e^{\im \frac{4\pi}{\beta}(\alpha-Q)}) \int_{\mathbb{R}} e^{   \left(\alpha-Q\right)c} \mathcal{G}^{\mathbb{S}}_{\beta}\left(\mu_B e^{  \frac{\beta}{2} c}\right) {\rm d}c.$$
Using the equation~\eqref{eq:laplace_integral-supp}, one ends up with the expression, with $s:=-\frac{2}{\beta}(\alpha-Q)$, 
\begin{align}
\left< V_{\alpha}(0)\right> &= \im (1-e^{\im \frac{4\pi}{\beta}(\alpha-Q)}) \int_{\mathbb{R}} \;e^{   \left(\alpha-Q\right)c}  \int _ 0 ^{ + \infty } e^{ -\overline{\mu}_Be^{\frac{\beta c}{2}} t ^{\frac{ \beta ^2 }{4}} - t} {\rm d}t  \,{\rm d}c\nonumber
\\
&= \im (1-e^{-2\pi s \im }) \frac{2}{\beta}\mu_B^{s}\Gamma(1+\tfrac{\beta^2}{4})^{-s}\Gamma(-s) \Gamma(1+\tfrac{\beta^2}{4}s)= \frac{4}{\beta} e^{-\im\pi s }\mu_B^{s}\frac{\Gamma(1+\tfrac{\beta^2}{4}s)}{(\Gamma(1+\tfrac{\beta^2}{4}))^{s}\Gamma(1+s)}\label{1pt-supp}
\end{align}

Note that we have the freedom to choose the normalization of the vertex operators. Of particular interest is the normalization making the normalized $1$-point function obey the two shift equations as well as the duality $\beta\to-4/\beta$.  More precisely, if we normalize the 
$1$-point function by $\langle U_{\alpha}(0)\rangle:=(e^{\im \pi \beta(\alpha-Q)}-1)\langle V_{\alpha}(0)\rangle$ then it obeys the two shift equations
\[\langle U_{\alpha+\frac{\beta}{2}}(0)\rangle=A\big(\frac{\beta}{2},\alpha,\mu\big)\langle U_{\alpha}(0)\rangle,\qquad \langle U_{\alpha-\frac{2}{\beta}}(0)\rangle=A\big(-\frac{2}{\beta},\alpha,\tilde\mu_B\big)\langle U_{\alpha}(0)\rangle\]
with the coefficient $A$ and the dual cosmological constant $ \tilde\mu_B$ given by
\[A(\chi,\alpha,\mu)=\frac{1}{\mu}\frac{1}{\chi^2} \frac{\sin(\pi\chi \alpha)}{\sin(\pi \chi (\alpha-Q))}\frac{\Gamma(1+\chi^2)\Gamma(-\chi\alpha)}{\Gamma(-\chi(\alpha-Q))},\qquad \tilde\mu_B:= (-\mu_B)^{-\frac{4}{\beta^2}}\Gamma(1+\tfrac{4}{\beta^2})    \Gamma(1+\tfrac{\beta^2}{4}) ^{\frac{4}{\beta^2}},\]
 as well as the duality relation $\langle U_{\alpha}(0)\rangle_{\mu_B,\beta}=\langle U_{\alpha}(0)\rangle_{\tilde\mu_B,-\frac{4}{\beta}}$. These computations show that the boundary LFT$_{\im \beta}$ 1-pt function features important CFT properties  based on the path integral \eqref{1-pt_path_integral}, and motivate further studies.  The crucial point is the decay of the Laplace transform \eqref{eq:momgen-supp}. Let us mention the case of the $2$-point function $\langle V_{\alpha_1}(0)V_{\alpha_2}(1)\rangle$ where the second insertion is chosen to be $\alpha_2=\frac{\beta}{2}$. The corresponding moment generating function writes $ \mathcal{G}^{\alpha_1,\frac{\beta}{2}}_{\beta}\left(\mu\right) = \E[e^{-\mu \int |e^{i\theta}-1|^{\frac{\beta^2}{2}} M_{\beta}^{\mathbb{S}}( d \theta)}]$ and its behaviour can be obtained by differentiating in $\mu$ the expression \eqref{eq:momgen-supp} and using rotational invariance. One gets that, up to a multiplicative factor $2\pi$, $ \mathcal{G}^{\alpha_1,\frac{\beta}{2}}_{\beta}\left(\mu\right) $ is the $\mu$-derivative of \eqref{eq:laplace_integral-supp}, which features an even better decay in $\mu$.

\section{LFT$_{\im \beta}$ on general surfaces and Seiberg bounds}

Let $(\Sigma,g)$ be a closed Riemannian surface and ${\bf x}:=\{x_1,\dots,x_n\}$ are $n$ marked points on $\Sigma$ with weights $\boldsymbol{\alpha}=(\alpha_1,\dots,\alpha_n)\in [Q,\infty)^n$, with $Q=\frac{\beta}{2}-\frac{2}{\beta}$ as before. Following the case of the disk, one is tempted to define the correlation functions by 
\[ \Big\langle \prod_{j=1}^nV_{\alpha_j}(x_j)\Big\rangle :=\lim_{\eps\to 0}\int_{\mathcal{C}} e^{isc} \; \E[\prod_{j=1}^n\eps^{\alpha_j^2/2}
e^{i\alpha_jX^{\Sigma}_{g,\eps}(x_j)}e^{\frac{i}{4\pi}Q\int K_gX_{g,\eps}{\rm dv}+\mu e^{i\beta c} \int_\Sigma\eps^{-\frac{\beta^2}{2}}e^{i\beta X_{g,\eps}}{\rm dv}}] \; {\rm d}c
\]
where $\mathcal{C}$ is a U-shaped contour as before, $s=\sum_{j=1}^n\alpha_j -Q\chi(\Sigma)$ with $\chi(\Sigma)$ the Euler characteristic, and $X_{g,\eps}$ the $\eps$-regularized GFF on $\Sigma$ with covariance given by the Green function of the Riemannian Laplacian $\Delta_g$. 
Using Cameron-Martin theorem we can shift the insertions $e^{i\alpha_jX_{g,\eps}(x_j)}$ and the curvature term to get 
\[
\begin{gathered} 
\Big\langle \prod_{j=1}^nV_{\alpha_j}(x_j)\Big\rangle:=F_{g,{\bf x},\boldsymbol{\alpha}}\int_{\mathcal{C}} e^{isc} \; \E[e^{\mu e^{i\beta c} 
M_{\beta,{\bf x},\boldsymbol{\alpha}}^{\Sigma}}]\; {\rm d}c,\\  
M_{\beta,{\bf x},\boldsymbol{\alpha}}^{\Sigma}:=\lim_{\epsilon\to 0}\int_\Sigma  e^{-\beta Q u(x)-\sum_j\beta\alpha_j G_g(x,x_j)}
\epsilon^{-\frac{\beta^2}{2}}e^{i\beta X_{g,\eps}(x)}{\rm dv}(x)\end{gathered}\]
with $u(x)=(4\pi)^{-1}\int_\Sigma G_g(x,x')K_g(x'){\rm dv}(x')$ and $F_{g,{\bf x},\boldsymbol{\alpha}}$ some explicit deterministic function involving $G_g$, $K_g$ and ${\bf x},\boldsymbol{\alpha}$. The probabilistic definition of the correlation functions can be proposed  by taking $\mathcal{C}$ to be the same contour as above, under the postulate that the Laplace transform $\E[e^{-\mu M_{\beta,{\bf x},\boldsymbol{\alpha}}^{\Sigma}}]$ decays with some power law $\mu^{-\lambda}$ as $\mu\to +\infty$ for some $\lambda>0$, provided the conformal weights satisfy 
\[ \alpha_j >Q ,\quad s=\sum_{j=1}^n\alpha_j -Q\chi(\Sigma)<\beta \lambda,\]
which can be viewed as \emph{Seiberg bounds} in that setting. If the Laplace transform has an asymptotic expansion of the form 
$\E[e^{-\mu M_{\beta,{\bf x},\boldsymbol{\alpha}}^{\Sigma,g}}]=C\mu^{-\lambda}+\mathcal{O}(\mu^{-\lambda-\epsilon})$ for $\epsilon>0$ as $\mu\to +\infty$ (as in the case of the circle above), we see that the correlation function has a pole at $s=\beta \lambda$.

\section{LFT$_{\im \beta}$ theory: numerical simulations}

We describe the implementation of our simulations for computing the multi-point correlation functions of the theory LFT$_{\im \beta}$ introduced in the main text. We focus on the choice of geometry and the range of validity of our simulations. To illustrate the method, we first present the simpler example of LFT$_{\im \beta}$ on the circle, for which the numerical results can be tested against exact predictions.
\subsection{Numerical simulation of LFT$_{\im \beta}$ on the circle}
Consider the unit disk of the complex plane, $\mathbb{D}$, equipped with the flat metric $g_0=|d z|^2$.  
The imaginary LFT$_{\im \beta}$ on a circle  is the LFT$_{\im \beta}$ on $\mathbb{D}$ with the bulk cosmological constant set to $0$, $\mu=0$, see Eq.~\eqref{imagLFT-supp}. Correlation functions involve an average over realizations of the (real) Neumann GFF $X_{g_0}$ defined by its covariance: $\mathbb{E} [ X_{g_0}(x)X_{g_0}(y)]  = - \log{|x-y||1-x\bar y|}$. The Neumann GFF can be split between contributions coming from the bulk and the boundary:
\begin{equation}
	X_{g_0}\stackrel{\text{law}}{=}X_{D,g_0}+\sqrt{2}P\varphi,
\end{equation}
where $X_{D, g_0}$ is the Dirichlet GFF on the unit disk and $P\varphi$ is the  harmonic extension inside the disk of $\varphi(\theta)$, the GFF on the unit circle $\mathbb{S}$ (independent of $X_{g_0,D}$). The covariance of the Dirichlet GFF is $\E[X_{D,g_0}(x)X_{D,g_0}(y)] = -\log |x-y| + \log |1-x\overline{y}|$. The boundary GFF has zero mean and covariance $\E [\varphi(\theta)\varphi(\theta^\prime) ]= -\log|e^{i\theta}-e^{i\theta'}|$.  The corresponding harmonic extension $P\varphi$ coincides with $\varphi$ on the boundary of the disk and has the covariance $\E[P\varphi(x)P\varphi(y)]=-\log|1-x\bar{y}|$ for $x,y\in\D$. $P\varphi$ can be decomposed on harmonic functions using the following spectral decomposition (or Karhunen-Loève expansion):
\begin{equation}
	P\varphi(z) = \sum_{n>0} \frac{a_n z^n}{\sqrt{2n}} + \sum_{n<0} \frac{\overline{a}_{-n} \overline{z}^{-n}}{\sqrt{-2n}} \ ,
	\label{eq:harmonic_extension}
\end{equation}
where $a_n$, $n>0$ are iid, complex normal distributed random variables.

In the section above, we derived exact results for the $1$-point function with one bulk insertion at the center of the disk ($x=0$). In that case, the only random field that enters the correlation functions is the boundary field $\varphi(\theta)=P\varphi(e^{\im \theta})$. 

\begin{figure}
	\begin{tikzpicture}
		\begin{axis}[
			axis x line=center,
			axis y line=center,
			xlabel={$\text{Re}(M^{\mathbb{S}}_{\beta,\epsilon})$},
			ylabel={$\text{Im}(M^{\mathbb{S}}_{\beta,\epsilon})$},
			xmin=-0.2,
			xmax=1.3,
			ymin=-0.1,
			ymax=0.1,
			]
			
			\addplot[only marks,mark size = 0.5pt,orange]  table {distMGCcircle.dat};
		\end{axis}
	\end{tikzpicture}
	
	\caption{We show the distribution of values of $M^{\mathbb{S}}_{\beta,\epsilon}$ for $\beta=0.43$}
	\label{data043}
\end{figure}
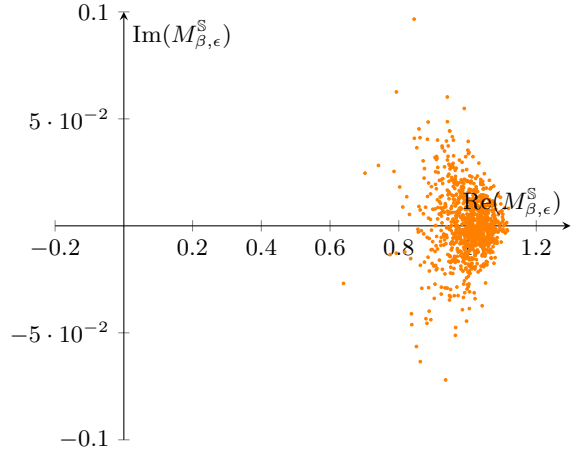

Numerically, we average over $N_{\text{sample}}$ samples of $\varphi_{\epsilon}(e^{\im \theta})$ of the boundary field:
\begin{equation}
	\varphi_{\epsilon}(\theta) = \sum_{n>0}^{N_{\text{modes}}} \frac{a_n e^{\im n \theta}}{\sqrt{2n}} + \sum_{n<0}^{N_{\text{modes}}} \frac{\overline{a}_{-n} e^{-\im n \theta}}{\sqrt{-2n}} \ .
\end{equation}
Each realization of the field $\varphi$ is constructed by randomly generating the random variables $a_{n}$ ($n=1,\cdots, N_{\text{modes}}$). The cutoff on the number of modes $N_\text{modes}$ acts as an UV regularization of the field. Roughly, it is equivalent to a spatial average of the GFF at scale $\epsilon = N_{\text{modes}}^{-1}$, since the variance of $\varphi_{\epsilon}$ scales as:
\begin{equation}
	\mathbb{E}\left[\varphi_{\epsilon}^2\right]\propto \log N_{\text{modes}} \propto -\log \epsilon.
\end{equation} 
We can now study the behavior of the random variable:
\begin{equation}
	M^{\mathbb{S}}_{\beta, \epsilon}= \frac{1}{2\pi} \int_{0}^{2\pi} e^{\frac{\beta^2}{4}\mathbb{E}\left[\varphi_{\epsilon}^2\right]}e^{\im \frac{\beta}{\sqrt{2}}\varphi_\epsilon(\theta)} {\rm d}\theta\ .
\end{equation}
Recall that the imaginary GMC $M^{\mathbb{S}}_{\beta}$~\eqref{imagLFT-supp} corresponds to the limit $M^{\mathbb{S}}_{\beta}=\lim_{\epsilon \to 0}M^{\mathbb{S}}_{\beta, \epsilon}$. In Fig.~\ref{data043}, we show  $N_{\text{sample}}=10^6$ instances of $M^{\mathbb{S}}_{\beta,\epsilon}$, generated with $\beta=0.43$ and $N_{\text{modes}}=100$.

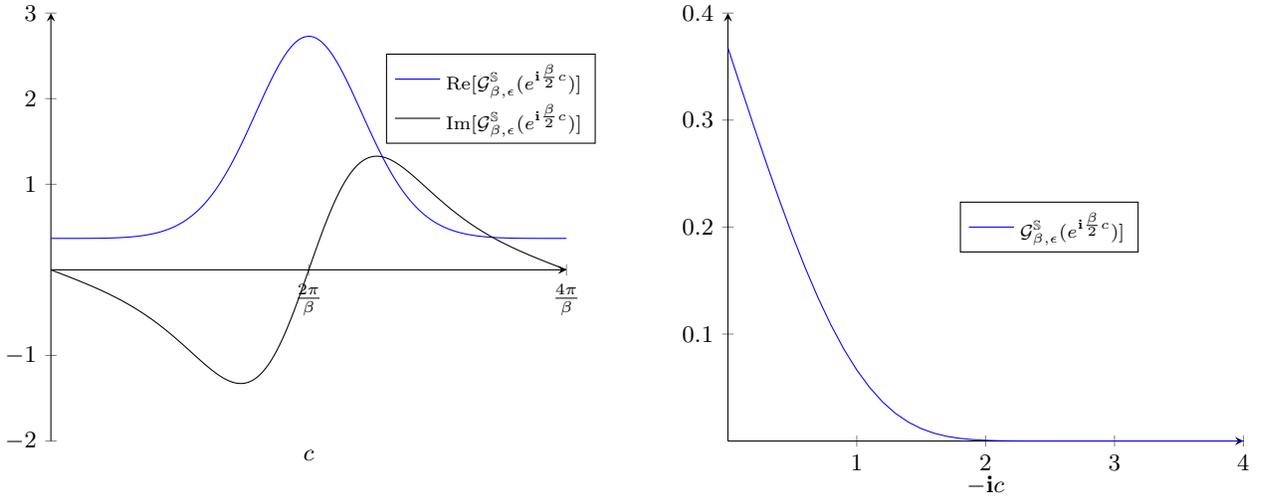
\begin{figure}
	\begin{tikzpicture}
		\begin{scope}
			\begin{axis}[
				axis x line=center,
				axis y line=center,
				xlabel={$c$},
				xlabel style={at={(axis description cs:0.5,0)}, anchor=north},
				xmin=0,
				xmax=6.28,
				ymin=-2,
				ymax=3,
				xtick={0,3.14,6.28},
				xticklabels={$0$,$\frac{2\pi}{\beta}$,$\frac{4\pi}{\beta}$},
				legend style={
					at={(0.65,0.8)},
					anchor=west,
					font=\scriptsize
				}
				]
				
				\addplot[blue]  table {Realvert.dat};
				\addlegendentry{$\text{Re}[\mathcal{G}^{\mathbb{S}}_{\beta,\epsilon}(e^{\im \frac{\beta}{2}c})]$}
				\addplot[red]  table {Imvert.dat};
				\addlegendentry{$\text{Im}[\mathcal{G}^{\mathbb{S}}_{\beta,\epsilon}(e^{\im \frac{\beta}{2}c})] $}
			\end{axis}
		\end{scope}
		
		\begin{scope}[xshift=9cm]
			\begin{axis}[
				axis x line=center,
				axis y line=center,
				xlabel={$-\im c$},
				xlabel style={at={(axis description cs:0.5,-0.06)}, anchor=north},
				xmin=0,
				xmax=4,
				ymin=0,
				ymax=0.4,
				legend style={
					at={(0.45,0.5)},
					anchor=west,
					font=\scriptsize
				}
				]
				
				\addplot[blue]  table {Horiz.dat};
				\addlegendentry{$\mathcal{G}^{\mathbb{S}}_{\beta,\epsilon}(e^{\im \frac{\beta}{2}c})] $}
				
			\end{axis}
		\end{scope}
		
	\end{tikzpicture}
	
	\caption{We show the behavior of $\mathcal{G}^{\mathbb{S}}_{\beta}(e^{\im \frac{\beta}{2}c})$ along the $\mathcal{U}$ contour, obtained by numerical simulations. On the left $c\in [0,\frac{4\pi}{\beta}]$, and on the right on the vertical line $c\in [0,-\im \infty]$.}
	\label{LT}
\end{figure}

Next, we turn on to the Laplace transform:
\begin{equation}
	\mathcal{G}^{\mathbb{S}}_{\beta,\epsilon}(\mu)=\mathbb{E}\big[e^{-\mu M^{\mathbb{S}}_{\beta,\epsilon}}\big] \ .
\end{equation} 
The results are shown in Fig.~\ref{LT}, where we investigate the behavior of $\mathcal{G}^{\mathbb{S}}_{\beta,\epsilon}(e^{\im \frac{\beta}{2}c})$ 
on the $\mathcal{U}$ contour defined above.

In Eq.~\eqref{circle_res1} or~\eqref{1pt-supp}, we gave a closed form formula for $\langle V_{\alpha}\rangle^{\text{LFT}_{\im \beta}}$ obtained by integrating the constant mode $c$ over the contour $\mathcal{U}$. 
From this formula, the function $\langle V_{\alpha}\rangle^{\text{LFT}_{   \im\beta }}$ vanishes for $s=\frac{2}{\beta}(Q-\alpha)\in-\mathbb{N}^*$, and has poles at  $s\in-\frac{4}{\beta^{2}}\mathbb{N}^*$. The presence of the poles is due to the algebraic decay of $\mathcal{G}^{\mathbb{S}}_{\beta}(\mu)$ for large positive $\mu$: $\mathcal{G}^{\mathbb{S}}_{\beta}(\mu)= \mathcal{O}(\mu^{-\frac{4}{\beta^{2}}}) $ as $\mu\to+\infty$ along the real line. Indeed, another term $e^{i\frac{\beta}{2}sc}$ enters the one-point correlation function~(\ref{1-pt_path_integral}), and compensates the decay of $\mathcal{G}^{\mathbb{S}}_{\beta}(\mu e^{i\frac{\beta}{2}sc } )$ precisely when $s=-\frac{4}{\beta^2}$. The other poles correspond to the values of $s$ matching the higher-order terms in the asymptotic expansion of the Laplace transform~(\ref{expansionG}).

We now compare (\ref{circle_res1}) to the numerical results. Since Eq.~\eqref{circle_res1} holds rigorously, the purpose of the comparison is to test the accuracy of the numerical method. The agreement is very good when $s$ is far from the first pole $s=-\frac{4}{\beta^2}$ (see Fig.~\ref{fincheck}). As a matter of fact, the agreement in this region stays strong even for a modest number of samples.
\begin{figure}
	\begin{tikzpicture}
		\begin{scope}
			\begin{axis}[
				axis x line=center,
				axis y line=center,
				xlabel={$\frac{2}{\beta}\left(\alpha-Q\right)$},
				xlabel style={at={(axis description cs:1,0.35)}, anchor=north},
				xmin=-4,
				xmax=4,
				ymin=-8,
				ymax=8,
				legend style={
					at={(0.75,1)},
					anchor=west,
					font=\scriptsize
				}
				]
				
				\addplot[only marks,mark size=1pt,orange]  table {Numre.dat};
				\addlegendentry{$\text{Re}[\left<V_{\alpha}\right>^{\text{LFT}_{\im \beta}}]$}
				
				\addplot[blue]  table {Thre.dat};
				\addlegendentry{Eq.~\eqref{circle_res1}}
			\end{axis}
		\end{scope}
		
		\begin{scope}[xshift=8cm]
			\begin{axis}[
				axis x line=center,
				axis y line=center,
				xlabel={$\frac{2}{\beta}\left(\alpha-Q\right)$},
				xlabel style={at={(axis description cs:1,0.35)}, anchor=north},
				xmin=-4,
				xmax=4,
				ymin=-8,
				ymax=8,
				legend style={
					at={(0.75,1)},
					anchor=west,
					font=\scriptsize
				}
				]
				
				\addplot[only marks,mark size=1pt,orange]  table {Numim.dat};
				\addlegendentry{$\text{Im}[\left<V_{\alpha}\right>^{\text{LFT}_{\im \beta}}]$}
				
				\addplot[blue]  table {Thim.dat};
				\addlegendentry{Eq.~\eqref{circle_res1}}
			\end{axis}
		\end{scope}

	\end{tikzpicture}
	
	\caption{The blue dots are the numerical results for $\left<V_{\alpha}\right>^{\text{LFT}_{\im \beta}}$, obtained by integrating over the $\mathcal{U}$ contour. The red curve is the analytic result, see Eq.~\eqref{circle_res1}. The numerical simulations depicted here correspond to $\beta=0.43$.}
	\label{fincheck}
\end{figure}
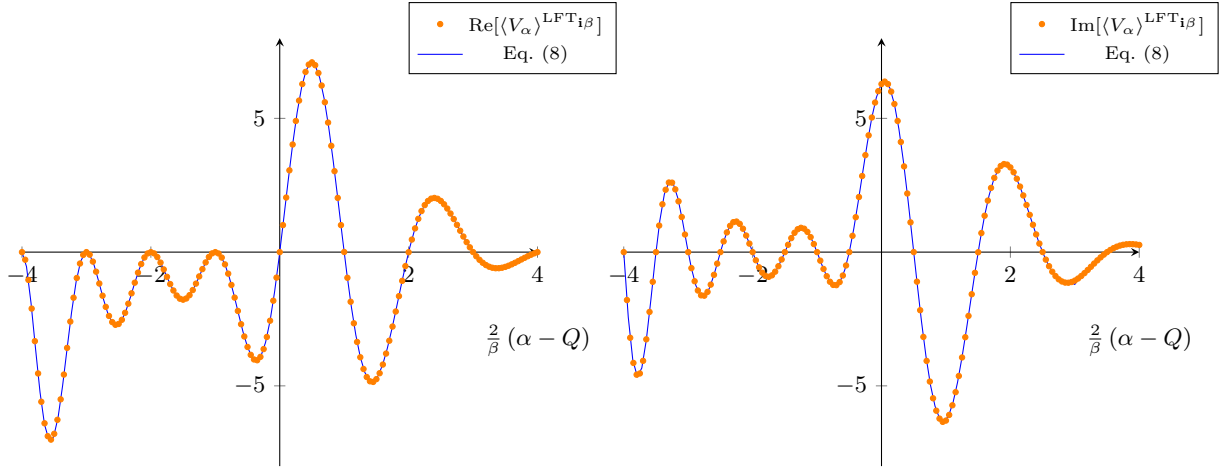

\begin{figure}
	\begin{tikzpicture}
		\begin{scope}
			\begin{axis}[
				axis x line=center,
				axis y line=center,
				xlabel={$s=\frac{2}{\beta}\left(\alpha-Q\right)$},
				xlabel style={at={(axis description cs:0.4,0)}, anchor=north},
				xmin=-22,
				xmax=4,
				ymin=-5,
				ymax=60,
				legend style={
					at={(0.12,0.85)},
					anchor=west,
					font=\scriptsize
				}
				]
				
				\addplot[only marks,mark size=0.5pt, orange]  table {Numlogre.dat};
				\addlegendentry{Numerical data}
				
				\addplot[blue]  table {Thlog.dat};
				\addlegendentry{$\log{\left|\langle V_{\alpha} \rangle^{\text{LFT}_{\im \beta}}\right|}$ from Eq.~(\ref{circle_res1})}
				
				\addplot[orange] coordinates {(-21.63,-10) (-21.63,60)};
				\addlegendentry{Pole at $s=-\frac{4}{\beta^2}$}
				
				\foreach \x in {-21,...,-1} {
					\addplot[blue, dashed] coordinates {({\x},-10) ({\x},60)};
				}
				\addlegendentry{Zeroes at $s\in-\mathbb{N}$}
				
			\end{axis}
		\end{scope}
		\begin{scope}[xshift=9cm]
			\begin{axis}[
				axis x line=center,
				axis y line=center,
				xmin=-21.4,
				xmax=-20,
				xlabel={$s=\frac{2}{\beta}\left(\alpha-Q\right)$},
				xlabel style={at={(axis description cs:0.4,1.15)}, anchor=north},
				legend style={
					at={(0.2,0.55)},
					anchor=west,
					font=\scriptsize
				}
				]
				
				\addplot[only marks,mark size=1pt, orange]  table {Numdet.dat};
				\addlegendentry{Numerical data}
				
				\addplot[blue]  table {Thdet.dat};
				\addlegendentry{$\text{Re}\left[\langle V_{\alpha} \rangle^{\text{LFT}_{\im \beta}}\right]$ from Eq.~(\ref{circle_res1})}
				
			\end{axis}
			
		\end{scope}    
	\end{tikzpicture}
	
	\caption{On the left diagram,  we plot the logarithm of $\left|\langle V_{\alpha}\rangle^{\text{LFT}_{\im \beta}}\right|$ in the domain $s\in [-\frac{4}{\beta^{2}}, 4]$, with $\beta=0.43$. On can observe that the numerical results (orange dots) perfectly match the analytic result for different order of magnitude. Approaching the pole at $s=-\frac{4}{\beta^{2}}$ the agreement is lost. On the right diagram represents a zoom on the region close to the pole where one clearly sees that numerical simulations fail to obtain the correct result.}
	\label{imposamp}
\end{figure}
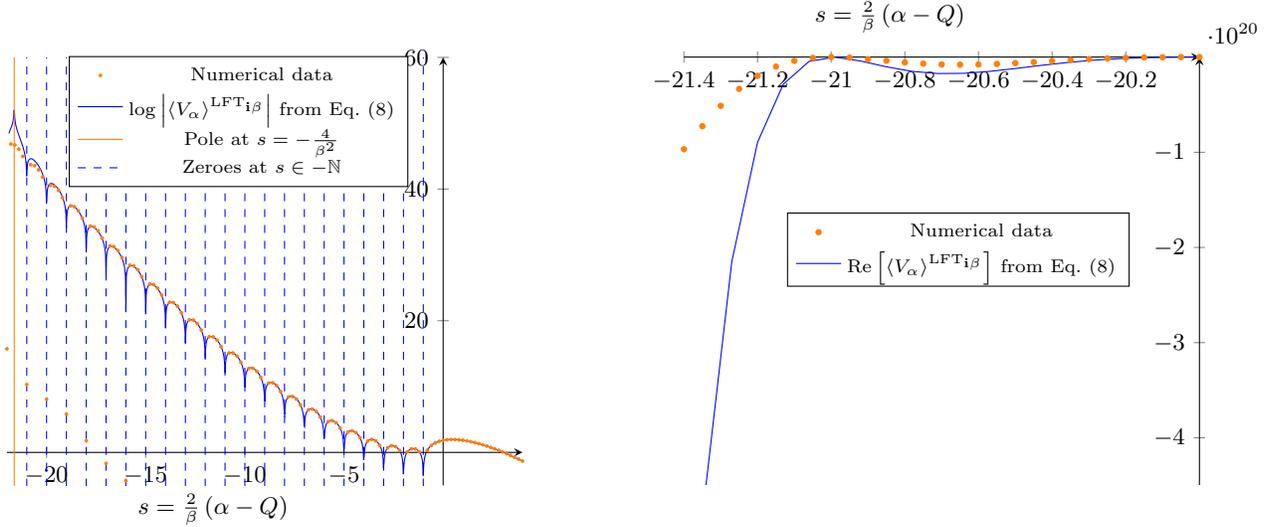
Nevertheless, the accuracy of the numerical results diminishes in the vicinity of the pole $s = -\frac{4}{\beta^2}$ (see Fig.~\ref{imposamp}). As mentioned earlier, the pole is determined by the asymptotic behavior of the Laplace transform $\mathcal{G}^{\mathbb{S}}_{\beta}(\mu)$ when $\mu\to+\infty$ along the real axis. In that region, the behavior of the Laplace transform is governed by the tail of the probability distribution of $M^{\mathbb{S}}_{\beta}$. It is highly plausible that our numerical simulations fail to adequately sample the very rare events that ultimately give rise to the pole. As we will see below, this is also a limitation of our simulations on the sphere.

\subsection{Numerical simulation of LFT$_{\im \beta}$ on the sphere}

We now turn to the three-point function of LFT$_{\im \beta}$ on the Riemann sphere. The main numerical bottleneck is the simulation of the Gaussian Free Field (GFF) on a two-dimensional surface. As we will see, our choice of geometry has two main advantages: i) we do not have to deal with finite size effects and ii) the background charge is localized on the unit circle and is very easy to implement. The price to pay is that the GFF is decomposed on a set of Bessel functions, which are numerically more expensive to evaluate than plane waves for which effective algorithms exist.

\subsubsection{Geometry setup}

To begin, we discuss the geometry used in the simulations. Again, let $\D$ be the unit disk in the complex plane $\C$ with flat metric $g_0=|dz|^2$, and let $\D^c = \mathbb{C}\backslash\mathbb{D}$ be equipped with the metric $g_0=|dz|^2/|z|^4$.
Let $X_1$ and $X_2$ be two independent Dirichlet GFFs on $\D$ and $\D^c$. They are defined by their covariances $\E [X_i(x)X_i(y)] = -\log |x-y| + \log |1-x\overline{y}|$ $(i=1,2)$. The Dirichlet GFF on the disk has the following decomposition on eigenfunctions of the Laplace-Beltrami operator:
\begin{equation}
	\label{DGFF}
	X_1(re^{i\theta}) = \sum_{n=1}^{\infty}\sum_{k=-\infty}^{\infty} a_{n,k} \frac{J_k(\alpha_{n,k}r) e^{ik\theta}}{\alpha_{n,k} J_{k+1}(\alpha_{n,k})}
\end{equation}
where $a_{n,k}$ are iid normal distributed random variables such that $a_{n,-k} = \overline{a_{n,k}}$. $J_k$ is the $k$-th Bessel function of first kind, and $\alpha_{n,k}$ is its $n$-th non-vanishing zero. Note that $(\D,g_0)$ and $(\D^c,g_0)$ are isometric \textit{via} $z\mapsto 1/\overline{z}$, and $X_2'(z):=X_2(1/\bar{z})\overset{\text{law}}{=}X_1(z)$.

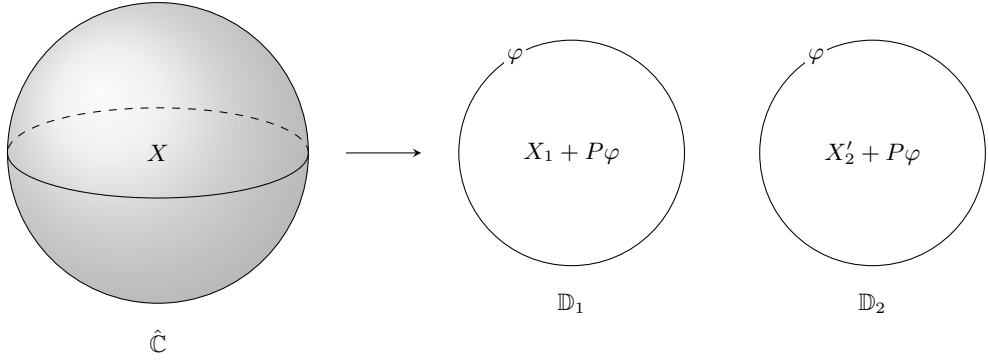
\begin{figure}
	\centering
	\begin{tikzpicture}
		\shade[ball color = gray!40, opacity = 0.4] (0,0) circle (2cm);
		\draw (0,0) circle (2cm);
		\draw (-2,0) arc (180:360:2 and 0.6);
		\draw[dashed] (2,0) arc (0:180:2 and 0.6);
		\node at (0,-2.5) {$\hat{\C}$};
		\node at (0,0) {\small $X$};
		\draw[-stealth] (2.5,0) -- (3.5,0);
		\node at (5.5,-2) { $\D_1$};
		\node at (9.5,-2) { $\D_2$};
		\draw (5.5,0) circle (1.5cm);
		\draw (9.5,0) circle (1.5cm);
		\node at (5.5,0) {\small $X_1+P\varphi$};
		\node at (9.5,0) {\small $X_2^\prime+P\varphi$};
		\node[fill=white,inner sep=1pt] at (4.75,1.29) {\small $\varphi$};
		\node[fill=white,inner sep=1pt] at (8.75,1.29) {\small $\varphi$};
	\end{tikzpicture}
	\caption{Identification of the Gaussian free field on the complex plane with fields on disks.}
	\label{fig:Markov}
\end{figure}

The GFF on the Riemann sphere is
\begin{equation}
	\label{eq:Markov}
	X(z) \overset{\text{law}}{=} \left\{\begin{array}{l}
		X_1(z) + P\varphi(z) \quad z\in\D \\
		X_2(z) + P\varphi(z) \quad z\in\D^c
	\end{array}\right.
\end{equation}
This Markov property of the GFF~\cite{sheffield2006gaussianfreefieldsmathematicians} is illustrated on Fig.~\ref{fig:Markov} (on which we also mapped $\D^c$ to the unit disk \textit{via} the isometry $z\to1/\overline{z}$). In our numerical simulations, the Liouville field comes in as the sum of the previous GFFs, plus the zero mode $c$.

\subsubsection{Definition of the observables}

We are interested in the $3$-point function $\langle  V_{\alpha_1} (x_1) V_{\alpha_2} (x_2) V_{\alpha_3} (x_3) \rangle$. By conformal symmetry, we can set $x_1=0$, $x_2=z\in \D$ and $x_3=\infty$. Define
\begin{equation}
	\label{eq:three_pt_amplitude}
	A_{{\bf x},\boldsymbol{\alpha},\epsilon}(c):= \E\Big[ \prod_{j=1}^3 \epsilon^{-\frac{ \alpha_j^2 } { 2 } }e^{i \alpha_j X_{\epsilon}(x_j)} e^{-\mu e^{i\beta c } \int_{\hat{\C}} \epsilon^{-\frac{\beta^2}{2}} e^{i \beta X_{\epsilon}(x) } {\rm dv}(x)} \Big] \ .
\end{equation}
$X_{\epsilon}$ is an ultraviolet regularization of $X$ defined by averaging on geodesic circles of radius $\epsilon$~\cite{guillarmou_polyakovs_2019}, or truncating over the number of modes in the spectral decompositions~\eqref{DGFF} and~\eqref{eq:harmonic_extension}. The three-point function is the integral
\begin{equation}
	\langle  V_{\alpha_1} (x_1) V_{\alpha_2} (x_2) V_{\alpha_3} (x_3) \rangle = \int_{\mathcal{U}} e^{i(\sum_{j=1}^3 \alpha_j - 2Q)c} A_{{\bf x},\boldsymbol{\alpha},\epsilon}(c) \; {\rm d}c \ .
\end{equation}

It is more convenient to use an alternative form of the amplitude~\eqref{eq:three_pt_amplitude}. Recall the \textit{Girsanov formula}~\footnote{ Also known by the name of \textit{complete-the-square trick} in the physics literature. }: if $Z$ is a centered Gaussian random variable and $F$ is a bounded functional over continuous functions, then
\begin{equation}
	\E \big[ e^{Z-\frac{1}{2}\E[Z^2]} F(X_\epsilon(x))\big] = \E \left[ F(X_\epsilon(x) + \E[Z X_\epsilon(x)])\right] \ .
\end{equation}
Note that we consider here the regularization $X_\epsilon$. Let us apply the Girsanov formula to $A_{{\bf x},\boldsymbol{\alpha},\epsilon}(c)$. We take successively $Z = \alpha_j X_{\epsilon}(x_j)$, $j=1,2,3$. The covariance of $X(x)$ is given by the Green function $G(x,y)=-\log|x-y|+\log|x|_++\log|y|_+ $, where $|x|_+=\max(1,|x|)$. The covariance of $X_{\epsilon}(x)$ is a smoothening of $G$ at scale $\epsilon$. We get:
\[ A_{{\bf x},\boldsymbol{\alpha},\epsilon}(c) = e^{-\sum_{j=1}^3\frac{\alpha^2_j}{2} m(x_j)-\sum_{i\neq j}\alpha_i\alpha_j G(x_i,x_j)} \E[ e^{-\mu e^{i\beta c } \int_{\hat{\C}} e^{- \sum_j \beta \alpha_j G(x,x_j)} \epsilon^{-\frac{\beta^2}{2}} e^{i\beta X_{\epsilon}(x)}{\rm dv}_g(x)} ] (1 + \underset{\epsilon\to0}{o}(1)) \]
where $m(x)=\lim_{x'\to x}G(x,x')+\log|x-x'|$. Using the Markov property~\eqref{eq:Markov}, we can now split the expectation as follows: $\E_{X} [\;\cdot\;] = \E_{\varphi} [ \E_{X_1} [\;\cdot\;] \E_{X_2}[\;\cdot\;] ]$ (where $\E_{Y}[\cdot]$ means that we integrate out the $Y$ random variable). We get:
\begin{subequations}
	\begin{align}
		A_{{\bf x},\boldsymbol{\alpha},\epsilon}(c) &= e^{\sum_{i\neq j}\alpha_i\alpha_j G(x_i,x_j)}  \E_{\varphi_\epsilon} [ A^1_{\epsilon}(c,\varphi)A^{2}_{\epsilon}(c,\varphi)] \nonumber \\
		A^1_{\epsilon}(c,\varphi)&:=\E_{X_{1,\epsilon}}[e^{-\mu e^{i\beta c} \int_{\D}|x|^{\beta\alpha_1}|x-z|^{\beta\alpha_2} \epsilon^{-\frac{\beta^2}{2}} e^{i\beta P\varphi_\epsilon(x)}e^{i\beta X_{1,\epsilon}(x)}dx}] \label{eq:amplitude_1} \\
		A^2_{\epsilon}(c,\varphi)&:=\E_{X_{2,\epsilon}^\prime}[e^{-\mu e^{i\beta c } \int_{\D}|x|^{\beta \alpha_3}\epsilon^{-\frac{\beta^2}{2}}e^{i\beta P\varphi_\epsilon(x)}e^{i\beta X_{2,\epsilon}^\prime(x)}dx}] \label{eq:amplitude_2}
	\end{align}
\end{subequations}
where in the last equality we used the fact that the integral over $\D^c$ can be expressed as an integral over $\D$ by the conformal change of variable $z\mapsto 1/\bar{z}$.

It is now clear that in a manner analogous to the standard Liouville theory~\cite{segal21}, the $3$-point function on the Riemann sphere is the result of gluing disk amplitudes:
\begin{equation}
	\label{eq:gluing}
	\langle  V_{\alpha_1} (0) V_{\alpha_2} (z) V_{\alpha_3} (\infty) \rangle = |z|^{-\alpha_1\alpha_2} \lim\limits_{\epsilon\to0} \int_{\mathcal{U}} e^{-i s \beta c} \; \E_{\varphi_\epsilon} [ A^{1}_{\epsilon}(c,\varphi) A^{2}_{\epsilon}(c,\varphi)]
\end{equation}
where $s = ( 2 Q -\sum_{j=1}^3 \alpha_j)/ \beta $.

\subsubsection{Numerical results}

We compare numerically the $3$-point function~\eqref{eq:gluing} and the imaginary DOZZ formula~\eqref{eq:Im_DOZZ}. This latter three-point function is an analytic continuation of the Coulomb-gas integrals when the charges do not fulfill the neutrality condition $s \in \N$.

\begin{figure}
	\includegraphics[width=.49\linewidth]{simu_dozz_24_04_v001.pdf}
	\includegraphics[width=.49\linewidth]{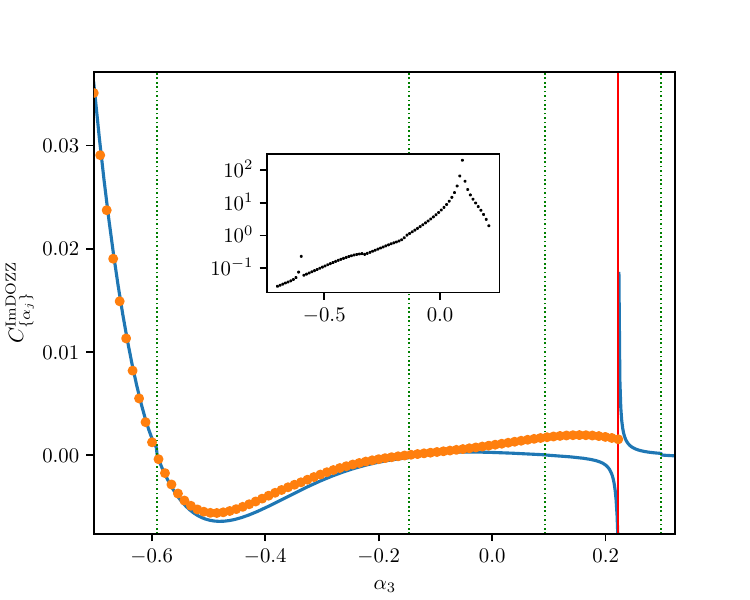}
	\caption{Comparison of the numerical results and the imaginary DOZZ formula (the values of the numerical parameters are $\beta=0.44$, $\alpha_1 = -3.94$, and $\alpha_2 = -3.58$, and the cutoff over modes of the GFF is $N_\text{modes}=2^{11}$). The area in the leftmost figure that is shaded in gray is out of reach of the probabilistic construction, due to the Seiberg bound $\alpha_3 > Q$ not being fulfilled. The poles and zeros of the imaginary DOZZ three-point function are marked with vertical lines in both figures. In the inserts, we show the relative deviation between the numerical data and the imaginary DOZZ. The deviation increases as we reach the pole $\alpha_3=\beta/2$ (red line on the rightmost figure).}
	\label{fig:match_overview-supp}
\end{figure}

We retrieve results similar to the case of Liouville theory on the circle (see Fig.~\ref{fig:match_overview-supp}). The match between numerical data and the imaginary DOZZ formula is very good across several orders of magnitude. Predictably, the error worsens when we reach the Seiberg bound $\alpha_3>Q$ from above. This comes from the divergence of the amplitudes~\eqref{eq:amplitude_1} and~\eqref{eq:amplitude_2}. More strikingly, the simulations lose accuracy when we come near to a pole of the imaginary DOZZ function. As before, we believe that it is due to insufficient sampling of the tails of the GMC distribution. We are currently investigating ways to improve our numerical method in this regard.

\bigskip
\noindent {\bf Acknowledgments.} We thank Xiangyu Cao for participating in the initial part of the work and for numerous subsequent discussions. We thank as well  Laurent Chevillard and Vincent Vargas for fruitful discussions and preliminary numerics in the early stages of this work. We acknowledge valuable discussions with Juhan Aru and Sylvain Ribault. R.~Usciati is supported by the CNRS 80 prime project. R.~Rhodes is partially supported by the Institut Universitaire de France (IUF) and by the ANR-21-CE40-0003. This work was performed using computational resources from the Mésocentre computing center of Université Paris-Saclay, CentraleSupélec and École Normale Supérieure Paris-Saclay supported by CNRS and Région Île-de-France.

\bibliographystyle{alpha}  
\bibliography{arxiv}

\end{document}